\documentclass[preprint,12pt]{elsarticle}
\usepackage{amssymb}
\usepackage{amsmath}
\usepackage{multirow}
\usepackage{colortbl}
\usepackage{xcolor}
\usepackage{rotating}
\usepackage{pgf}
\usepackage{array}
\usepackage{booktabs}
\usepackage{pifont}
\usepackage{amssymb}
\usepackage{graphicx}
\usepackage{subcaption}
\usepackage{setspace}
\usepackage{amsmath}
\usepackage{nccmath}
\usepackage{float}
\usepackage{framed}

\usepackage{hyperref}

\hypersetup{
    colorlinks=true,        
    linkcolor=black,        
    citecolor=blue,         
    urlcolor=gray,          
    filecolor=magenta,      
}

\journal{Computer Speech \& Languages}

\begin{document}

\begin{frontmatter}

\title{Causal Analysis of ASR Errors for Children:\\
Quantifying the Impact of Physiological, Cognitive, and Extrinsic Factors}

\author[label1]{Vishwanath Pratap Singh}
 \affiliation[label1]{organization={University of Eastern Finland},
             addressline={Tulliportinkatu 1},
             city={Joensuu},
             postcode={80130},
             country={Finland}}

\author[label2,label3]{Md. Sahidullah}
\affiliation[label2]{organization={Institute for Advancing Intelligence, TCG CREST},
             city={Kolkata},
             postcode={700091},
             country={India}}

\affiliation[label3]{organization={Academy of Scientific and Innovative Research (AcSIR)},
             city={Ghaziabad},
             postcode={201002},
             country={India}}

\author[label1]{Tomi H. Kinnunen}


\begin{abstract}
The increasing use of children’s automatic speech recognition (ASR) systems has spurred research efforts to improve the accuracy of models designed for children's speech in recent years. The current approach utilizes either open-source speech foundation models (SFMs) directly or fine-tuning them with children’s speech data. These SFMs, whether open-source or fine-tuned for children, often exhibit higher word error rates (WERs) compared to adult speech. However, there is a lack of systemic analysis of the cause of this degraded performance of SFMs.  Understanding and addressing the reasons behind this performance disparity is crucial for improving the accuracy of SFMs for children's speech. Our study addresses this gap by investigating the causes of accuracy degradation and the primary contributors to WER in children’s speech. In the first part of the study, we conduct a comprehensive benchmarking study on two self-supervised SFMs (\texttt{Wav2Vec 2.0} and \texttt{Hubert}) and two weakly supervised SFMs (\texttt{Whisper} and \texttt{MMS}) across various age groups on two children speech corpora, establishing the raw data for the causal inference analysis in the second part. In the second part of the study, we analyze the impact of physiological factors (age, gender), cognitive factors (pronunciation ability), and external factors (vocabulary difficulty, background noise, and word count) on SFM accuracy in children's speech using causal inference. The results indicate that physiology (age) and particular external factor (number of words in audio) have the highest impact on accuracy, followed by background noise and pronunciation ability. Fine-tuning SFMs on children’s speech reduces sensitivity to physiological and cognitive factors, while sensitivity to the number of words in audio persists.  
\end{abstract}

\begin{graphicalabstract}
\end{graphicalabstract}

\begin{highlights}
\item A causal framework to analyze factors contributing to ASR errors for children.

\item We include physiological, cognitive, and extrinsic factors in our analysis.

\item Two self-supervised (\texttt{Wav2Vec2.0} and \texttt{Hubert}) and two weakly supervised  (\texttt{Whisper} and \texttt{MMS}) SFMs in analysis. 

\item Insights into fine-tuning of speech foundation models.

\end{highlights}


\end{frontmatter}



\section{Introduction}
\label{intro}
Children of the twenty-first century are frequently referred to as \emph{digi-natives} due to their upbringing in an era where digital technology is ubiquitous and readily accessible~\cite{diginative}. Consequently, they constitute one of the primary demographics for daily usage of smartphones, tablets, and other digital devices~\cite{oecd}. This increased digital proficiency has catalyzed a surge in research on speech technology for children, including the task of transcribing speech into text, i.e. \emph{automatic speech recognition} (ASR)~\cite{gurunath_csl,slt_child}. 
However, ASR models' efficacy for children remains (by a large margin) below those designed for adults~\cite{child_patel,lpc_aug}. Addressing this shortcoming necessitates a deeper understanding of the factors contributing to the accuracy gap and the formulation of strategies to mitigate the identified challenges.

Similar to other speech tasks, modern children's ASR leverages from \emph{speech foundation models} (SFMs)~\cite{sfm,elderly} pre-trained on diverse and large quantities of speech data to learn generalizable representation space usable in various speech tasks. SFMs can be utilized directly as pre-trained open-source models~\cite{child_patel,jain_w2v} or be further fine-tuned to children's speech data 
While our present study is not the first one to quantify the impact of factors responsible for ASR performance degradation for children, we address several methodological shortcomings of prior research. Previous studies, such as~\cite{analysis_icslp}, have investigated the impact of children's pronunciation abilities on 
classic statistical (pre deep learning era) ASR systems within a small classroom setup and a fixed age group. Even if more recent work \cite{gurunath_csl,analysis_small_age_groups} has addressed the impact of age and the number of words in context on substitution, deletion, and insertion errors on modern deep learning era ASR, 
the studies 
typically 
analyze explaining factors 
in isolation, i.e. overlooking their interdependent effects. 
To see why this can be problematic, consider the joint development of physiological characteristics and cognitive skills in children. Both of them depend on \emph{age}---the possibly most studied performance variable in children ASR. 
Nonetheless, age is a potential \emph{confounder} 
which can mask the more subtle, actual cause of ASR errors. 

In general, when commonplace correlational methods are used to analyze data that includes confounders, there is a 
risk to draw erroneous conclusions. A known shortcoming of traditional correlational (aka, \emph{associative} or \emph{statistical}) approaches is that they analyze co-variation between the variables included (e.g. age vs. WER)---but do not explicitly differentiate between causes and effects. Causality, however, is arguably among the fundamental building blocks for refining any scientific theory or model. By incorporating relevant background knowledge to  random variables and their assumed causal relations, causal methodologies aim at findings with stronger generalization potential, compared to (purely) correlational findings. Compare this, for instance, to well-established equations in physics (like Newton's second law) that are arguably more universal compared to a linear regressor fitted to a random data sample but without explicit understanding of the mechanisms that give rise to the observations. Even if causal methodologies are rarely used in the speech field, causality itself is well-matured \cite{pearl2009}. We hope the present study, in the context of children speech processing, to serve as possible inspiration for further research in the field.

Besides the complex relation of physiological and cognitive factors that underlie speech production, \emph{extrinsic} factors (ranging from the characteristics of the acoustic environment to the complexity of the text material) are expected to influence ASR errors---again, however, these factors are rarely included in the error analysis. Finally, the selected SFM model (a ``machine listener'') itself influences the ASR error patterns. In fact, any model contains built-in \emph{inductive bias} which is impacted by the properties of the training data, the model architecture, and the training loss~\cite{gurunath_csl} to name a few. To account for the risks related to selections of specific ASR models and datasets,  
we conduct experiments using 
four modern ASR models that are representative of both 
self- and weakly-supervised 
models. 
To account for the risks related to choice of datasets, we include two different datasets---CSLU Kids~\cite{cslu} and MyST~\cite{myst}---in our experiments. To go beyond the default metric of ASR, WER, we also analyze the substitution, deletion, and insertion errors.


To sum up, there is a clear need for a generalizable causal analysis framework that takes these various considerations into account holistically and help to pave way towards addressing challenges that underlay children ASR. 
Understanding the delicate interdependencies of the various potential contributing factors is essential for accurately characterizing the underlying causes of ASR degradation. This understanding may help to guide the design of novel features, model architectures, and datasets. 
To the best of our knowledge, our study marks the first known effort of its kind. With the help of the proposed framework, we seek to answer the following, more specific research questions:

\begin{enumerate} 
\item \textbf{What is the relative impact of physiological, cognitive, and external factors on the performance of open-source SFMs for children's speech?} 

\item \textbf{Pronunciation of children's impacts the WER~\cite{analysis_icslp}. But, what influences children’s pronunciation ability more: their age or the complexity of the vocabulary they use?} 

\item \textbf{How does fine-tuning speech foundation models on child-specific speech datasets reduce the model's sensitivity to children's age, pronunciation ability, and extrinsic factors? Which aspect benefits the most from fine-tuning, and which remains largely unaffected?}

 \end{enumerate}

The remainder of the paper is organized as follows. After situating our study to the broader landscape of explainability and causality (Section \ref{sec:relwork}), the main experimental part follows in the next two Sections. Section \ref{case1} establishes the necessary, more traditional analysis of ASR errors. This section provides insights beyond WER, a detailed analysis of SFM model complementarities, including both their off-the-shelf and fine-tuned use cases). The causal modeling approach, detailed Section \ref{case2}, details the variables (not used in prior studies) and their specifical causal structure, in addition to reporting the inference of causal strength between the variables through both linear and nonlinear approach, detailed below. The big-picture take-home messages and provided through the two last sections of Discussion and Conclusion.

\section{Related Work}
\label{sec:relwork}

Methodologically, our study is situated within the broader field of \emph{explainability}. 
As such, explainability is central to scientific knowledge discovery aimed at improving one's understanding of a given phenomenon. In the scope of artificial intelligence and machine learning, explainability is described as the degree to which a human can understand the cause of a decision~\cite{miller2017}, or the ability to explain in ways that are intelligible to a human~\cite{doshi2017}. Explainability 
approaches are generally categorized into two main groups. 
(1) The first one, \emph{inherently explainable} models, 
generates explanations during training \cite{yang2016}. 
These models generate explanations directly through their structure, enabling straightforward understanding. Canonical examples include \emph{decision trees} that represent their decisions through hierarchical ``\texttt{if-then}'' rules; 
\emph{linear regression} that can be used for quantifying 
feature importance through analysis of the learned weights; and \emph{attention networks} that highlight important input features~\cite{bahdanau2014, attention, yang2016}. 
(2) The second, 
\emph{post-hoc} category of explainability approaches, provides explanations for decisions \emph{after} they have been made \cite{kim2016, renkl2014}. 
Examples for post-hoc explanations include \emph{local interpretable model-agnostic explanations} (LIME), which approximates complex black-box models locally~\cite{ribeiro2016} (applied for 
ASR analysis in~\cite{LIME-TIMIT}), \emph{saliency maps} that identify influential pixels in machine vision tasks 
~\cite{simonyan2014}, and \emph{feature visualization} that shows what neural networks learn via activation maximization~\cite{erhan2009}.

A unifying methodology framework that covers both inherent and post-hoc explainable models is \textbf{causal inference} (CI)~\cite{explainable-ml}. CI 
considers hidden or unobserved factors influencing decision-making and enabling counterfactual explanations~\cite{explainable-ml}. 
Historically, CI methods have been used more in the clinical domain, where understanding the underlying mechanisms of disease progression and determining the impacts of treatment effects are clearly of paramount importance. This includes explainable decision-making processes in data collection, annotation, and pre-processing~\cite{causal-medical}. More recently, causality-aided machine learning has also been addressed through \emph{causal representation learning} 
~\cite{causal-rpl}. Overall, the CI mindset, targeted at understanding the underlying cause-effect mechanisms, has been 
less common in machine learning, including research in speech technology. Although machine learning and CI share common \emph{tools} (e.g., graphical models and latent variables), the sentiment in the former is to optimize predictive performance, as opposed to interpreting the parameters to learn about causal relations, as in the latter. In a sense, predictive performance and explainability can be conflicting goals: heavily over-parameterized deep neural network models (including ASR) apparently generalize well, whereas the individual parameters (nowadays billions of network weights) are hardly meaningful to us, as humans.

Our aim in the present study is to provide a post-hoc explainable analysis of the general error mechanisms underlying children's ASR, addressed through the current state-of-the-art SFMs. While ASR technology has always been a fast-moving target (and dependent on the availability of data, compute, interests of researchers among other factors), arguably the physiological and cognitive factors that underlie children language learning, including their speech production development, are more fundamental. In order to understand better the current challenges, as well as pave way for future ASR development for children, it is important to 
be able to quantify 
how the physiological, cognitive, and extrinsic factors 
interact and jointly contribute to 
ASR performance. Here, CI 
provides a natural analysis framework. 
Accordingly, we employ CI to generate post-hoc explanations for SFMs tailored to children's speech, enabling a deeper understanding of the factors contributing to the degraded word error rate (WER). 

Causality provides a 
principled framework for recovering and estimating cause and effect variables from data observations. Any causality problem can be divided into two consecutive sub-problems of \emph{causal structure discovery} and \emph{causal inference} \cite{gendron2023}. Whereas the former aims at recovering the structure of causal relations, in the form of a \emph{directed acyclic graph} (DAG), the latter estimates the functions that link 
the nodes in a given DAG. The causal structure can either be learned automatically \cite{colombo2012, spirtes1991, spirtes2000}, or be constructed from prior human knowledge \cite{zhou2024}. Once the causal structure has been established, CI can be used to estimate the links between variables. While the classic CI approaches include \emph{do-calculus}~\cite{pearl2009}, Bayesian inference frameworks, and regression methods \cite{gelman2006}, 
recent approaches 
(e.g. \cite{xia2021}) also leverage from deep neural networks---
particularly, the graph neural network \cite{zecevic2021}---to estimate causal dependency functions. 
We provide a summary of the causal structure discovery and quantification methods being used for post-hoc explanation of machine learning models in Table~\ref{tab:ci}.

\begin{table}[h!]
\centering
\tiny
\caption{Summary of Causal methods being used in post-hoc explainability in machine learning.}
\begin{tabular}{|c|c|c|c|}
\hline
\textbf{Publication} & \textbf{Task} & \textbf{Causal Structure Discovery Method} & \textbf{Causal Quantification Method} \\
\hline
\cite{animal} & Animal Behavioral Modeling  & PCMI~\cite{PCMI} & Graph Neural Network \\ 
\hline
\cite{ci_reco} & Explainability in Recommendation System& Domain Knowledge & Logistic Regression \\
\hline
\cite{zhou2024} & Sentiment Classification in NLP & Domain Knowledge & Bayesian Network\\
\hline

Our & Causes of degradation in children's ASR & Domain Knowledge & Bayesian Network \\
\hline
\end{tabular}
\label{tab:ci}
\end{table}

For simplicity and to leverage existing domain knowledge, we construct our 
DAG based on domain knowledge presented in~\cite{workforce_children}. Then, we use the Bayesian inference framework~\cite{pearl2009} to estimate the link between nodes in the causal graph. The strength of the causal relationship between the cause and the effect nodes in DAG can then be quantified using either the \emph{average causal effect} (ACE) or \emph{conditional mutual information} (CMI)~\cite{causal-quantification,causal-survey}. While ACE captures linear cause-effect relationships, CMI accounts for non-linear dependencies between cause and effect. We point the interested reader to Appendix~\ref{app1} for a more detailed explanation of causal inference, and to~\cite{pearl} for further details. 

\begin{table}[t]
 \tiny
 \vspace{-0.05cm}
  \caption{CSLU kids Corpus (scripted): \# speakers and \# of utterances from different genders in the scripted set of the CSLU kids corpus. This protocol was developed in our previous study~\cite{childaugment}.}
 \centering
 \begin{tabular}{|c | c | c | c |c | c| c|} 
 \hline
 \multicolumn{1}{|c|}{} &\multicolumn{3}{c|}{Boys} & \multicolumn{3}{c|}{Girls}\\
\cline{2-7}
 Split & \# Spk & \# Utt  & {\# Hours} & \# Spk & \# Utt & {\# Hours}\\ 
 \hline\hline
Original Dataset & 605 & 39184 & {38.0} & 513 & 32815& {31.8} \\
\hline
Cleaned &  602 & 39167 & {37.9} & 511 & 32811 & {31.8}\\
\hline
With \textbf{good} tag & 602 & 25454 &  {22.6} &	511 &	22071 &	{19.7} \\ 
 \hline
Test-good & 542 & 22960 & {20.4} & 451 & 19561 & {17.6}\\
 \hline
Dev-good & 30 & 1290 & {1.1} & 30 & 1260 & {1.1}\\ 
\hline
Train-good & 30 & 1204 & {1.1} & 30 & 1250 & {1.1}\\ 
 \hline
 \end{tabular}
 \label{data}
\end{table} w2ws

\begin{figure}[!t]
    \centering
     \includegraphics[width=0.7\textwidth]{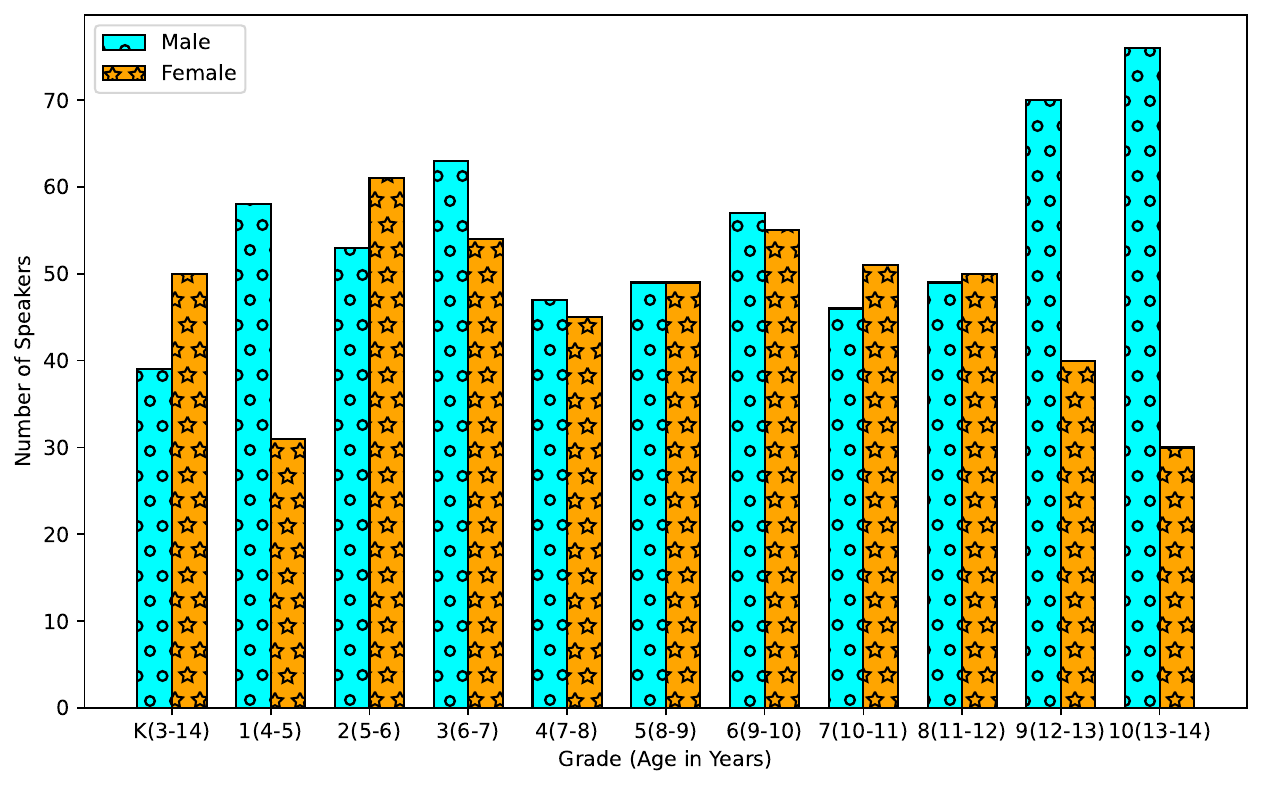}
     \caption{Gender-wise speaker distribution per grade (age in years) group in the CSLU Kids corpus.}
     \label{fig:cslu}
 \end{figure}

\setlength{\tabcolsep}{3pt}
\begin{table}[!t]
    \tiny
    \centering
    \caption{Comparison of speech foundation models (SFMs) in terms of architecture, loss, training data, and the number of parameters.}
    \begin{tabular}{|c|c|c|c|c|c|}
        \hline
        \textbf{Model} & \textbf{Architecture} & \textbf{Training Strategy} & \textbf{Pretraining Data} & \textbf{Training Data} & \textbf{\# Params} \\
        \hline
        \texttt{Wav2Vec2.0-large} & Encoder only & Self-supervised &  Libri-light & Libri-960 & 317M (large) \\
        \hline
        \texttt{HuBERT-large} & Encoder only & Self-supervised loss & Libri-light& Libri-960 & 317M million (base) \\
        \hline
        \texttt{Whisper-base} & Encoder-Decoder & Weakly-supervised & - & Unknown 680K Hours & 244M (small) \\
        \hline
        \texttt{MMS-1.1B} & Encoder only  & Weakly-supervised  & - & Unknown & 300M \\
        \hline
    \end{tabular}
    \label{tab:sfms}
\end{table}

\begin{table}[!t]
\renewcommand{\arraystretch}{1.1}
\centering
\caption{Comparison of overall ASR errors on MyST and CSLU Kids with open-source SFMs and fine-tuned SFMs. Subs, Del, and Ins indicate substitution, deletion, and insertion errors (in \%). We only fine-tune the best open-source self-supervised, and best open-source weakly supervised SFMS. Hence, the result corresponding to fine-tuned HuBERT and MMS models remains empty. }
\vspace{0.2cm}
\tiny
\begin{tabular}{|c|c|c|c|c|c|c|c|}
\hline
\textbf{Model} & \textbf{Error} & \multicolumn{3}{c|}{\textbf{Open Source}} & \multicolumn{3}{c|}{\textbf{Fine-Tuned}} \\
\cline{3-8}
 & & \textbf{MyST Dev} & \textbf{MyST Test} & \textbf{CSLU Read} & \textbf{MyST Dev} & \textbf{MyST Test} & \textbf{CSLU Read} \\
\hline
\multirow{4}{*}{Wav2Vec} & WER & 21.6 & 25.2 & 54.0 & 13.2 & 15.6 & 32.4 \\
 & Subs & 9.0 & 10.8 & 21.6 & 6.0 & 7.2 & 14.4 \\
 & Del & 5.4 & 5.4 & 10.8 & 3.6 & 3.6 & 7.2 \\
 & Ins & 7.2 & 9.0 & 21.6 & 4.8 & 6.0 & 14.4 \\
\hline
\multirow{4}{*}{HuBERT} & WER & 19.8 & 23.4 & 48.6 &-  & - & - \\
 & Subs & 7.2 & 9.0 & 16.2 &- &- &  -\\
 & Del & 3.6 & 5.4 & 10.8 & -& - & -\\
 & Ins & 9.0 & 10.8 & 27.0 & - & -&  -\\
\hline
\multirow{4}{*}{Whisper} & WER & 18.0 & 21.6 & 26.28 & 18.23 & 19.71 & 25.2 \\
 & Subs & 5.4 & 7.2 & 16.95 & 3.6 & 4.8 & 10.8 \\
 & Del & 3.6 & 3.6 & 3.64 & 2.4 & 2.4 & 3.6 \\
 & Ins & 7.2 & 9.0 & 6.39 & 4.8 & 6.0 & 14.4 \\
\hline
\multirow{4}{*}{MMS} & WER & 23.4 & 27.0 & 59.4 & - & -&  -\\
 & Subs & 10.8 & 12.6 & 27.0 & -& -&  -\\
 & Del & 7.2 & 7.2 & 16.2 & - &- & - \\
 & Ins & 9.0 & 10.8 & 27.0 & - & -&  -\\
\hline
\end{tabular}
\label{tab:res1}
\end{table}

\begin{table}[!t]
\tiny
 \caption{Detailed WER analysis across gender and age groups on CSLU Test set using open-source and fine-tuned speech foundation models. OS indicates the open source models, while FT indicates the models fine-tuned on MyST train split. Subs, Del, and Ins indicate substitution, deletion, and insertion errors (in \%). Oracle Hypo. Selection results are after selecting the best hypothesis (best on WER) of an utterance from four models.}
 \label{tab:res2}
  \vspace{-0.0 cm}
 \centering
 \begin{tabular}{|c | c | c | c |c |c |c|c|c|c|c|c|c|c|c|c|c} 
 \cline{4-14}
 \multicolumn{3}{ c| }{} &\multicolumn{11}{c|}{Grades}\\
\cline{2-14}

\multicolumn{1}{ c| }{} & \multicolumn{1}{ c| }{Model} &  \multicolumn{1}{ c| }{(\%)} &\multicolumn{1}{ c| }{K} & \multicolumn{1}{ c| }{1} & \multicolumn{1}{ c| }{2} & \multicolumn{1}{ c| }{3}& \multicolumn{1}{ c| }{4} & \multicolumn{1}{ c| }{5} & \multicolumn{1}{ c| }{6} & \multicolumn{1}{ c| }{7}& \multicolumn{1}{ c| }{8} & \multicolumn{1}{ c| }{9} & \multicolumn{1}{ c| }{10} \\ 
 \cline{2-14}\hline
\multirow{20}{*}{OS} & \multirow{4}{*}{wav2vec} & WER & \cellcolor{red!66.8}66.8 & \cellcolor{red!42.4}42.4 & \cellcolor{red!36.1}36.1 & \cellcolor{red!30.4}30.4 & \cellcolor{red!23.8}23.8 & \cellcolor{red!22.3}22.3 & \cellcolor{red!64.6}64.6 & \cellcolor{red!56.8}56.8 & \cellcolor{red!55.0}55.0 & \cellcolor{red!56.9}56.9 & \cellcolor{red!55.6}55.6 \\
& & Subs & \cellcolor{blue!46.8}46.8 & \cellcolor{blue!30.3}30.3 & \cellcolor{blue!26.4}26.4 & \cellcolor{blue!21.3}21.3 & \cellcolor{blue!17.2}17.2 & \cellcolor{blue!15.8}15.8 & \cellcolor{blue!46.2}46.2 & \cellcolor{blue!40.3}40.3 & \cellcolor{blue!39.1}39.1 & \cellcolor{blue!39.9}39.9 & \cellcolor{blue!38.1}38.1 \\
& & Del & \cellcolor{yellow!7.6}7.6 & \cellcolor{yellow!3.3}3.3 & \cellcolor{yellow!2.5}2.5 & \cellcolor{yellow!1.3}1.3 & \cellcolor{yellow!0.9}0.9 & \cellcolor{yellow!1.3}1.3 & \cellcolor{yellow!0.0}0.0 & \cellcolor{yellow!0.0}0.0 & \cellcolor{yellow!0.0}0.0 & \cellcolor{yellow!0.0}0.0 & \cellcolor{yellow!0.0}0.0 \\
& & Ins & \cellcolor{green!12.4}12.4 & \cellcolor{green!8.8}8.8 & \cellcolor{green!7.2}7.2 & \cellcolor{green!7.9}7.9 & \cellcolor{green!5.7}5.7 & \cellcolor{green!5.2}5.2 & \cellcolor{green!18.4}18.4 & \cellcolor{green!16.5}16.5 & \cellcolor{green!15.9}15.9 & \cellcolor{green!17.0}17.0 & \cellcolor{green!17.5}17.5 \\
  \cline{3-14}
   & \multirow{4}{*}{HuBert} & WER & \cellcolor{red!71.6}71.6 & \cellcolor{red!42.5}42.5 & \cellcolor{red!34.6}34.6 & \cellcolor{red!29.5}29.5 & \cellcolor{red!21.0}21.0 & \cellcolor{red!20.8}20.8 & \cellcolor{red!53.3}53.3 & \cellcolor{red!45.8}45.8 & \cellcolor{red!44.1}44.1 & \cellcolor{red!49.2}49.2 & \cellcolor{red!45.5}45.5 \\
& & Subs & \cellcolor{blue!46.9}46.9 & \cellcolor{blue!28.9}28.9 & \cellcolor{blue!23.6}23.6 & \cellcolor{blue!19.0}19.0 & \cellcolor{blue!14.8}14.8 & \cellcolor{blue!14.5}14.5 & \cellcolor{blue!37.9}37.9 & \cellcolor{blue!33.6}33.6 & \cellcolor{blue!32.0}32.0 & \cellcolor{blue!34.6}34.6 & \cellcolor{blue!31.1}31.1 \\
& & Del & \cellcolor{yellow!4.9}4.9 & \cellcolor{yellow!2.6}2.6 & \cellcolor{yellow!1.8}1.8 & \cellcolor{yellow!1.0}1.0 & \cellcolor{yellow!0.7}0.7 & \cellcolor{yellow!1.1}1.1 & \cellcolor{yellow!0.0}0.0 & \cellcolor{yellow!0.0}0.0 & \cellcolor{yellow!0.0}0.0 & \cellcolor{yellow!0.0}0.0 & \cellcolor{yellow!0.0}0.0 \\
& & Ins & \cellcolor{green!19.9}19.9 & \cellcolor{green!11.0}11.0 & \cellcolor{green!9.1}9.1 & \cellcolor{green!9.5}9.5 & \cellcolor{green!5.6}5.6 & \cellcolor{green!5.2}5.2 & \cellcolor{green!15.3}15.3 & \cellcolor{green!12.2}12.2 & \cellcolor{green!12.1}12.1 & \cellcolor{green!14.6}14.6 & \cellcolor{green!14.4}14.4 \\
  \cline{3-14}

 & \multirow{4}{*}{whisper} & WER& \cellcolor{red!84.6}84.6& \cellcolor{red!37.5}37.5& \cellcolor{red!28.0}28.0& \cellcolor{red!23.1}23.1& \cellcolor{red!17.5}17.5& \cellcolor{red!18.7}18.7& \cellcolor{red!30.3}30.3& \cellcolor{red!26.3}26.3& \cellcolor{red!22.0}22.0& \cellcolor{red!23.1}23.1& \cellcolor{red!22.4}22.4
 \\
& & Subs & \cellcolor{blue!34.8}34.8& \cellcolor{blue!22.4}22.4& \cellcolor{blue!18.0}18.0& \cellcolor{blue!14.1}14.1& \cellcolor{blue!10.4}10.4& \cellcolor{blue!11.3}11.3& \cellcolor{blue!21.3}21.3& \cellcolor{blue!18.5}18.5& \cellcolor{blue!16.9}16.9& \cellcolor{blue!16.5}16.5& \cellcolor{blue!16.1}16.1
 \\
& & Del & \cellcolor{yellow!2.4}2.4& \cellcolor{yellow!4.9}4.9& \cellcolor{yellow!4.0}4.0& \cellcolor{yellow!4.2}4.2& \cellcolor{yellow!4.7}4.7& \cellcolor{yellow!4.6}4.6& \cellcolor{yellow!0.0}0.0& \cellcolor{yellow!0.0}0.0& \cellcolor{yellow!0.0}0.0& \cellcolor{yellow!0.0}0.0& \cellcolor{yellow!0.0}0.0 \\
& & Ins & \cellcolor{green!47.4}47.4& \cellcolor{green!10.2}10.2& \cellcolor{green!6.1}6.1& \cellcolor{green!4.8}4.8& \cellcolor{green!2.4}2.4& \cellcolor{green!2.7}2.7& \cellcolor{green!9.0}9.0& \cellcolor{green!7.8}7.8& \cellcolor{green!5.2}5.2& \cellcolor{green!6.6}6.6& \cellcolor{green!6.3}6.3 \\
  \cline{3-14}
   & \multirow{4}{*}{MMS} & WER & \cellcolor{red!100}122.4 & \cellcolor{red!85.8}85.8 & \cellcolor{red!87.6}87.6 & \cellcolor{red!68.4}68.4 & \cellcolor{red!60.0}60.0 & \cellcolor{red!57.3}57.3 & \cellcolor{red!100}193.3 & \cellcolor{red!100}168.3 & \cellcolor{red!100}131.5 & \cellcolor{red!100}166.6 & \cellcolor{red!100}150.9 \\
& & Subs & \cellcolor{blue!62.0}62.0 & \cellcolor{blue!46.2}46.2 & \cellcolor{blue!43.5}43.5 & \cellcolor{blue!36.7}36.7 & \cellcolor{blue!32.5}32.5 & \cellcolor{blue!31.4}31.4 & \cellcolor{blue!78.3}78.3 & \cellcolor{blue!73.2}73.2 & \cellcolor{blue!70.5}70.5 & \cellcolor{blue!76.4}76.4 & \cellcolor{blue!71.1}71.1 \\
& & Del & \cellcolor{yellow!7.6}7.6 & \cellcolor{yellow!4.4}4.4 & \cellcolor{yellow!4.0}4.0 & \cellcolor{yellow!2.1}2.1 & \cellcolor{yellow!1.7}1.7 & \cellcolor{yellow!2.1}2.1 & \cellcolor{yellow!0.0}0.0 & \cellcolor{yellow!0.0}0.0 & \cellcolor{yellow!0.0}0.0 & \cellcolor{yellow!0.0}0.0 & \cellcolor{yellow!0.0}0.0 \\
& & Ins & \cellcolor{green!52.7}52.7 & \cellcolor{green!35.2}35.2 & \cellcolor{green!40.2}40.2 & \cellcolor{green!29.5}29.5 & \cellcolor{green!25.8}25.8 & \cellcolor{green!23.8}23.8 & \cellcolor{green!100.0}115.0 & \cellcolor{green!95.0}95.0 & \cellcolor{green!61.0}61.0 & \cellcolor{green!90.1}90.1 & \cellcolor{green!79.8}79.8 \\
  \cline{3-14}

   & \multirow{4}{*}{Oracle Hypo. Selection} & WER & \cellcolor{red!42.47}42.47 & \cellcolor{red!27.34}27.34 & \cellcolor{red!21.33}21.33 & \cellcolor{red!18.26}18.26 & \cellcolor{red!14.18}14.18 & \cellcolor{red!15.47}15.47 & \cellcolor{red!23.63}23.63 & \cellcolor{red!20.13}20.13 & \cellcolor{red!17.83}17.83 & \cellcolor{red!17.62}17.62 & \cellcolor{red!17.60}17.60 \\
& & Subs & \cellcolor{blue!30.18}30.18 & \cellcolor{blue!18.47}18.47 & \cellcolor{blue!14.41}14.41 & \cellcolor{blue!11.59}11.59 & \cellcolor{blue!8.57}8.57 & \cellcolor{blue!9.62}9.62 & \cellcolor{blue!17.46}17.46 & \cellcolor{blue!15.28}15.28 & \cellcolor{blue!13.81}13.81 & \cellcolor{blue!13.19}13.19 & \cellcolor{blue!13.09}13.09 \\
& & Del & \cellcolor{yellow!2.19}2.19 & \cellcolor{yellow!4.00}4.00 & \cellcolor{yellow!2.90}2.90 & \cellcolor{yellow!3.23}3.23 & \cellcolor{yellow!4.16}4.16 & \cellcolor{yellow!3.98}3.98 & \cellcolor{yellow!0.00}0.00 & \cellcolor{yellow!0.00}0.00 & \cellcolor{yellow!0.00}0.00 & \cellcolor{yellow!0.00}0.00 & \cellcolor{yellow!0.00}0.00 \\
& & Ins & \cellcolor{green!10.09}10.09 & \cellcolor{green!4.87}4.87 & \cellcolor{green!4.02}4.02 & \cellcolor{green!3.44}3.44 & \cellcolor{green!1.45}1.45 & \cellcolor{green!1.86}1.86 & \cellcolor{green!6.17}6.17 & \cellcolor{green!4.85}4.85 & \cellcolor{green!4.02}4.02 & \cellcolor{green!4.43}4.43 & \cellcolor{green!4.51}4.51 \\
 \hline
 \hline

\multirow{12}{*}{FT} &  \multirow{4}{*}{wav2vec} & WER & \cellcolor{red!88.1}88.1& \cellcolor{red!62.2}62.2& \cellcolor{red!58.3}58.3& \cellcolor{red!52.0}52.0& \cellcolor{red!45.7}45.7& \cellcolor{red!50.5}50.5& \cellcolor{red!100}107.9& \cellcolor{red!100}109.6& \cellcolor{red!100}112.0& \cellcolor{red!100}106.6& \cellcolor{red!92.8}92.8 \\
& & Subs  & \cellcolor{blue!54.3}54.3& \cellcolor{blue!41.3}41.3& \cellcolor{blue!39.0}39.0& \cellcolor{blue!35.2}35.2& \cellcolor{blue!31.3}31.3& \cellcolor{blue!33.8}33.8& \cellcolor{blue!64.2}64.2& \cellcolor{blue!65.5}65.5& \cellcolor{blue!67.4}67.4& \cellcolor{blue!63.9}63.9& \cellcolor{blue!59.8}59.8
\\
& & Del& \cellcolor{yellow!6.9}6.9& \cellcolor{yellow!2.4}2.4& \cellcolor{yellow!2.1}2.1& \cellcolor{yellow!1.5}1.5& \cellcolor{yellow!1.1}1.1& \cellcolor{yellow!1.7}1.7& \cellcolor{yellow!0.0}0.0& \cellcolor{yellow!0.0}0.0& \cellcolor{yellow!0.0}0.0& \cellcolor{yellow!0.0}0.0& \cellcolor{yellow!0.0}0.0
 \\
& & Ins & \cellcolor{green!26.9}26.9& \cellcolor{green!18.5}18.5& \cellcolor{green!17.2}17.2& \cellcolor{green!15.2}15.2& \cellcolor{green!13.3}13.3& \cellcolor{green!14.9}14.9& \cellcolor{green!43.7}43.7& \cellcolor{green!44.1}44.1& \cellcolor{green!44.6}44.6& \cellcolor{green!42.7}42.7& \cellcolor{green!32.9}32.9 \\
  \cline{3-14}
 & \multirow{4}{*}{whisper} & WER & \cellcolor{red!62.8}62.8& \cellcolor{red!45.9}45.9& \cellcolor{red!40.7}40.7& \cellcolor{red!43.6}43.6& \cellcolor{red!31.7}31.7& \cellcolor{red!33.4}33.4& \cellcolor{red!57.2}57.2& \cellcolor{red!51.5}51.5& \cellcolor{red!54.2}54.2& \cellcolor{red!53.4}53.4& \cellcolor{red!47.2}47.2 \\
& & Subs & \cellcolor{blue!38.9}38.9& \cellcolor{blue!26.3}26.3& \cellcolor{blue!23.7}23.7& \cellcolor{blue!20.9}20.9& \cellcolor{blue!17.7}17.7& \cellcolor{blue!18.3}18.3& \cellcolor{blue!37.6}37.6& \cellcolor{blue!34.9}34.9& \cellcolor{blue!36.1}36.1& \cellcolor{blue!33.7}33.7& \cellcolor{blue!32.2}32.2 \\
& & Del & \cellcolor{yellow!3.3}3.3& \cellcolor{yellow!7.1}7.1& \cellcolor{yellow!6.5}6.5& \cellcolor{yellow!6.7}6.7& \cellcolor{yellow!6.3}6.3& \cellcolor{yellow!6.7}6.7& \cellcolor{yellow!0.0}0.0& \cellcolor{yellow!0.0}0.0& \cellcolor{yellow!0.0}0.0& \cellcolor{yellow!0.0}0.0& \cellcolor{yellow!0.0}0.0 \\
& & Ins & \cellcolor{green!20.6}20.6& \cellcolor{green!12.6}12.6& \cellcolor{green!10.5}10.5& \cellcolor{green!16.0}16.0& \cellcolor{green!7.7}7.7& \cellcolor{green!8.4}8.4& \cellcolor{green!19.6}19.6& \cellcolor{green!16.7}16.7& \cellcolor{green!18.1}18.1& \cellcolor{green!19.7}19.7& \cellcolor{green!15.0}15.0 \\
  \cline{3-14}

   & \multirow{4}{*}{Oracle Hypo. Selection} & WER & \cellcolor{red!36.52}36.52 & \cellcolor{red!23.51}23.51 & \cellcolor{red!18.35}18.35 & \cellcolor{red!15.70}15.70 & \cellcolor{red!12.20}12.20 & \cellcolor{red!13.31}13.31 & \cellcolor{red!20.32}20.32 & \cellcolor{red!17.31}17.31 & \cellcolor{red!15.32}15.32 & \cellcolor{red!15.15}15.15 & \cellcolor{red!15.14}15.14 \\
& & Subs & \cellcolor{blue!25.34}25.34 & \cellcolor{blue!15.89}15.89 & \cellcolor{blue!12.39}12.39 & \cellcolor{blue!9.96}9.96 & \cellcolor{blue!7.37}7.37 & \cellcolor{blue!8.28}8.28 & \cellcolor{blue!15.21}15.21 & \cellcolor{blue!13.14}13.14 & \cellcolor{blue!11.88}11.88 & \cellcolor{blue!11.30}11.30 & \cellcolor{blue!11.22}11.22 \\
& & Del & \cellcolor{yellow!1.89}1.89 & \cellcolor{yellow!3.44}3.44 & \cellcolor{yellow!2.49}2.49 & \cellcolor{yellow!2.78}2.78 & \cellcolor{yellow!3.58}3.58 & \cellcolor{yellow!3.42}3.42 & \cellcolor{yellow!0.00}0.00 & \cellcolor{yellow!0.00}0.00 & \cellcolor{yellow!0.00}0.00 & \cellcolor{yellow!0.00}0.00 & \cellcolor{yellow!0.00}0.00 \\
& & Ins & \cellcolor{green!9.29}9.29 & \cellcolor{green!4.18}4.18 & \cellcolor{green!3.47}3.47 & \cellcolor{green!3.08}3.08 & \cellcolor{green!1.25}1.25 & \cellcolor{green!1.61}1.61 & \cellcolor{green!5.11}5.11 & \cellcolor{green!4.17}4.17 & \cellcolor{green!3.44}3.44 & \cellcolor{green!3.85}3.85 & \cellcolor{green!3.92}3.92 \\
  \cline{3-14}
 \hline
 \end{tabular}
 \vspace{-0.05cm} 
\end{table}

\begin{figure}[t]
\centering
\includegraphics[width=\textwidth]{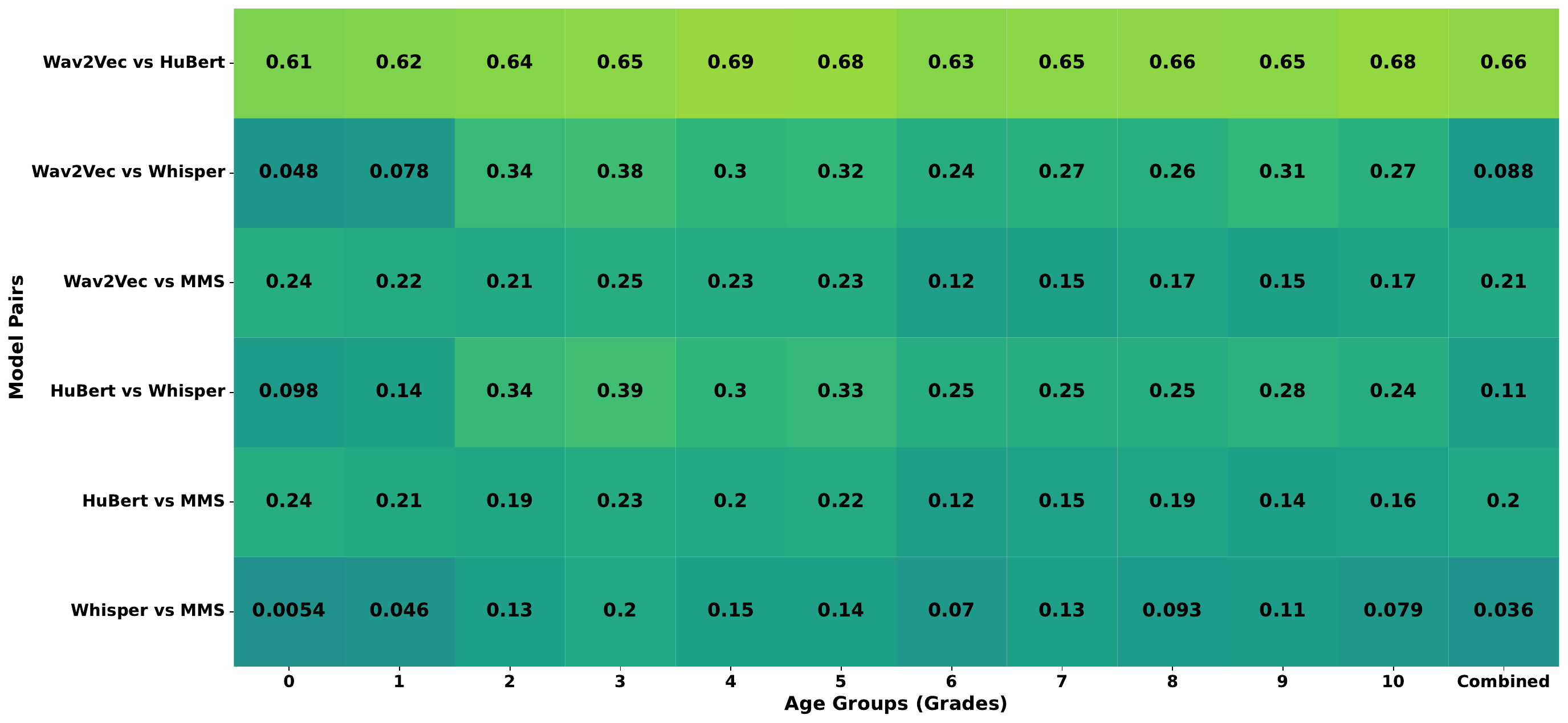}
\caption{Correlation heat maps between different open-source speech foundation models across various age groups of children in the CSLU Test corpus.}
\label{fig:cor2}
\end{figure}

\section{Analysis of ASR Errors}
\label{case1}
The causal post-hoc modeling experiments presented in Section \ref{case2} requires ASR models to be constructed. This section serves for this purpose, producing the raw data needed for causal modeling. 

\subsection{Experimental Setup}
\subsubsection{Dataset}
We consider two standard children speech corpora in our experiments: CSLU Kids~\cite{cslu} and My Science Tutor (MyST) \cite{myst}. The 
former comprises both spontaneous and scripted recordings from approximately 1100 children. It includes approximately a gender-balanced group of roughly 100 children per grade, spanning from kindergarten (age 5-6 years old) to grade 10 (age 15-16 years old). Gender-wise speaker distribution per age group is shown in Figure \ref{fig:cslu}. 
Following~\cite{child_benchmark}, we focus exclusively on the scripted subset in the CSLU kids corpus. It includes isolated words and sentences from 605 boys and 513 girls, categorized as ``Good'' (only target word), ``Maybe'' (target word with extra content), ``Bad'' (no target word), and ``Puff'' (target word with inhaling noise). For our experiments, we include only the ``Good'' recordings. This gives a customized, speaker-disjoint train, development, and test sets for 
summarized in Table \ref{data}. The age and gender metadata 
enables SFM performance evaluation across age and gender groups.

The second and more recent corpus, MyST, 
consists of 470 hours of English speech from 1371 students in grades 3 to 5 engaged in conversations with a tutor. The corpus includes speaker-disjoint subsets for training (1052 speakers), development (153 speakers), and testing (156 speakers). Some of the MyST recordings are longer than 30 seconds, where self-supervised models such as \texttt{Hubert} and \texttt{Wav2Vec2.0} may fail as they are trained on audio lengths of 0.5 seconds to 30 seconds. Hence, we removed recordings in excess of 30 seconds. 
Different from CSLU Kids, the age and gender metadata are not available for the MyST dataset, which restricts us from presenting 
age- and gender-specific results on MyST. 

\subsubsection{Automatic Speech Recognition Systems}
As summarized in Table~\ref{tab:sfms}, we consider four 
ASR systems in our experiments: \texttt{Wav2Vec 2.0}, \texttt{HuBERT}, \texttt{Whisper}, and \texttt{MMS}. 
The first one, \texttt{Wav2Vec2.0}~\cite{wav2vec}, is a self-supervised model, which leverages contrastive~\cite{closs} training to learn contextually informed speech representations from unlabeled audio data. The model comprises a pre-trained transformer encoder and a quantized feature encoder, enabling efficient and effective representation learning from raw audio. The second model, \texttt{HuBERT}~\cite{hubert}, is also self-supervised, 
based on a similar architecture to \texttt{Wav2Vec2.0}. Different from \texttt{Wav2Vec2.0}, however, \texttt{HuBERT} applies a classification task, forcing the model to classify hidden sequences into pre-defined categories. 
Finally, the two last models, 
\texttt{Whisper}~\cite{whisper} and \texttt{MMS}~\cite{mms}, 
are \emph{weakly supervised} models that rely on limited labeled data and large amount of unlabeled data for training. While \texttt{Whisper} utilizes a semi-supervised approach to learn from both labeled and unlabeled data, 
\texttt{MMS} adopts a weakly supervised multi-task learning framework to jointly optimize for various speech processing tasks. In the first part of the analysis, we utilize these open-source SFMs `as-is', while in the second part, we address the impact of fine-tuning on best performing models. 


\subsection{Results and Discussion}
We present the overall 
WER, substitution, deletion, and insertion errors for the open-source models and their fine-tuned 
variants 
in Table~\ref{tab:res1}. 
\texttt{Whisper}~\cite{whisper} outperforms the other three models, 
indicating its generalizability, which might be due to a large number of parameters, inclusion of diverse training data, or both. Furthermore, we note that the substitution errors are consistently higher compared to the two other types of errors across the models and test sets. 
A further breakdown of the results according to 
age and 
gender 
is provided in Table~\ref{tab:res2}. 
Again, \texttt{Whisper} 
outperforms the other models, while 
\texttt{MMS}~\cite{mms} yields the highest WERs. We observe an overall decreasing trends in all 
the metrics with increasing grade (a proxy for age), following 
trends reported in other studies~\cite{child_patel,jain_w2v,gurunath_csl,jain_whisper}. 
The apparent `jump' between grades 5 and 6 
can be attributed to the use of different sentences in grades K to 5 and 6 to 10. The latter consists primarily of single-word utterances, 
which results in notably high WERs. 
A higher WER for the single word utterances can be attributed to the architecture limitation of SFMs: it is known that models trained with attention-based architectures~\cite{attention} benefit from the larger context size, which is less common in the upper grades, where shorter utterances prevail. Additionally, 
the two self-supervised models, 
\texttt{Wav2Vec2.0}~\cite{wav2vec} and \texttt{HuBERT}~\cite{hubert}, exhibit a relatively larger increase in errors between grades 5 and 6, indicating their increased sensitivity to shorter context. This might be attributed to the use of 
\emph{contrastive loss}~\cite{closs}, which requires large context to predict the masked input representation. Finally, as for gender, 
we observe no notable change in WER trends 
between boys and girls. 

Overall, the performance of the four ASR models is far from perfect, with notably high error rates for the younger age groups and the for single-word utterances. It is therefore interesting to consider the combined potential of the four models under an oracle selection strategy. In this ``oracle hypothesis selection'' strategy, summarized in the last group of results in Table~\ref{tab:res2}, we output the transcript from whichever model yields the lowest WER \emph{on the specific utterance}. This is an oracle experiment aimed at providing insights into 
the complementarity of the different SFMs at the level of 
individual utterances. 
As one may expect, this reduces WERs substantially, 
suggesting that the four models produce differing transcripts for the same utterance in the CSLU Kids corpus. Since this analysis does not reveal \emph{which} groups of models are more/less complementary to each other, 
we 
further report the correlation coefficient between each model pair on the utterance-level WER data. The results, summarized in Fig.~\ref{fig:cor2}, indicate that \texttt{Wav2Vec} and \texttt{HuBERT} are most strongly correlated. This 
might be due to the similarity in the architectures and training datasets of these two models; both are trained in a self-supervised manner on LibriSpeech~\cite{libri}. Interestingly, \texttt{Whisper} shows the lowest correlation with the remainder of the models, which might be due to its distinct training data and architecture.

Up to this point, we have focused on generic open-source models whose checkpoints (neural network weights) are publicly available. None of the four models are particularly optimized for children ASR. While the above \emph{zero-shot} evaluation of ASR on an unseen target domain (here, childrens' speech) is justified from a practical deployment point of view, whenever speech data \emph{is} available from the target domain, it is beneficial to employ model fine-tuning~\cite{jain_w2v,gurunath_csl,jain_whisper}. For this experiment, based on the above zero-shot results, we select \texttt{Wav2Vec2.0} and \texttt{Whisper} to represent self- and weakly-supervised model families, respectively. 
These two models are fine-tuned on the MyST train split, the MyST dev data being used to select the best checkpoint. We do not use CSLU Train-good (defined in Table ~\ref{data}) as it is too small for training (compared to MyST train split). The overall results 
are presented in Table~\ref{tab:res1}, with the breakdown by age being presented in Table~\ref{tab:res2}. 

While the fine-tuned models show a reduced WER for on both 
MyST Test and CSLU Test-good, the relative reduction in the former is greater. 
This might be due to the diverse age groups in the CSLU Kids corpus, whereas the age distribution in the MyST training and test sets is limited to grades 5 to 8. Furthermore, the age-wise results of the fine-tuned models in Table~\ref{tab:res2} are better than those of the open-source models, but they show a similar trend of WER reduction as the grade level increases. Similar to open source models, 
we see the `jump' between grades 5 and 6, 
indicating that fine-tuning 
did not reduce the sensitivity to shorter contexts. 

\begin{figure}[!t]
    \centering
\tiny
    \begin{subfigure}[t]{0.85\textwidth}
        \centering
        \resizebox{\textwidth}{!}{
        \begin{tabular}{|c|c|c|c|}
            \hline
            \mathtt{\textbf{Type~of~Response}} & \mathtt{\textbf{Modeling~Variable}} & \mathtt{\textbf{Available in Metadata/Inferred}} & \mathtt{\textbf{Node~Type~in~Causal~Graph}} \\ 
            \hline
            \multirow{2}{*}{\mathtt{Physiological}} & \mathtt{Age} & \mathtt{Available} & \mathtt{Confounder} \\
            \cline{2-4}
                                         & \mathtt{Gender} & \mathtt{Available} & \mathtt{Fork} \\
            \hline
            \mathtt{Cognitive} & \mathtt{GoP} & \mathtt{Inferred} & \mathtt{Collider} \\
            \hline
            \multirow{3}{*}{\mathtt{Extrinsic}} & \mathtt{SNR} & \mathtt{Inferred} & \mathtt{Fork} \\
            \cline{2-4}
                                     & \mathtt{Vocab~Difficulty} & \mathtt{Inferred} & \mathtt{Fork} \\
            \cline{2-4}
                                     & \mathtt{\#Words~in~Audio} & \mathtt{Available} & \mathtt{Fork} \\
            \hline
        \end{tabular}
        }
        \caption{}
        \label{fig:table-var}
    \end{subfigure}

\vspace{0.2cm}
    \begin{subfigure}[t]{0.48\textwidth}
        \centering
        \includegraphics[width=0.8\textwidth]{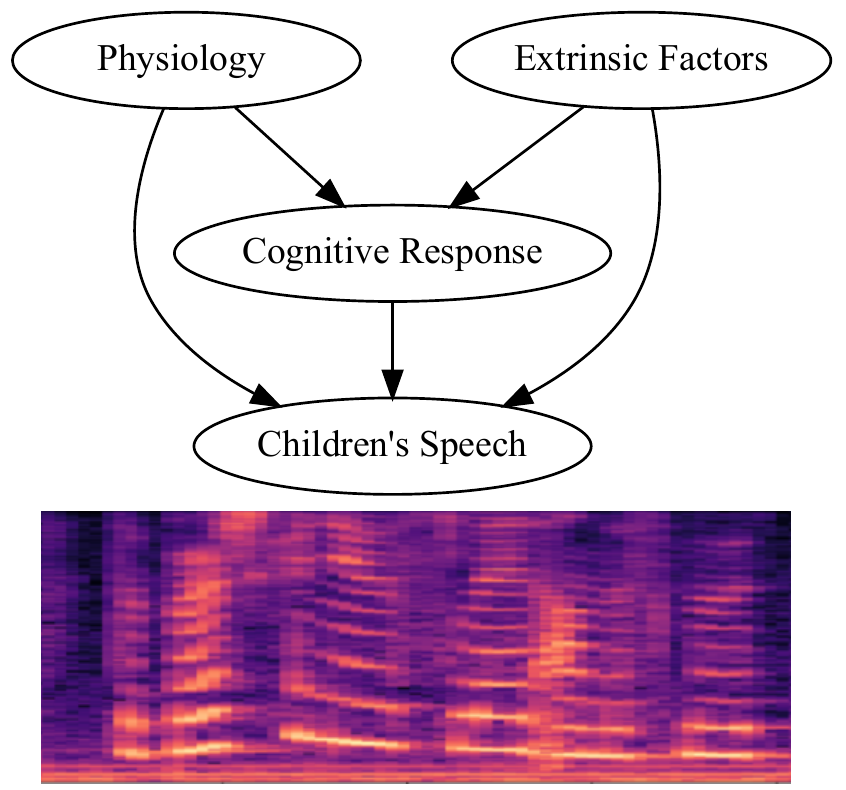}
        \caption{}
        \label{fig:speech}
    \end{subfigure}
\hfill
\begin{subfigure}[t]{0.48\textwidth}
    \centering
    \vspace{-3.5cm}
    \begin{equation*}
        \vspace{0.1cm}
        \begin{flalign*}
        &\mathtt{Rule~1:~Age} \rightarrow \{\mathtt{Subs,~Del,~Ins~Error}\} & \\
        &\mathtt{Rule~2:~Gender} \rightarrow \{\mathtt{Subs,~Del,~Ins~Error}\} & \\
        &\mathtt{Rule~3:~Vocab~Diff} \rightarrow \{\mathtt{Subs,~Del,~Ins~Error}\} & \\
        &\mathtt{Rule~4:~GoP} \rightarrow \{\mathtt{Subs,~Del,~Ins~Error}\} & \\
        &\mathtt{Rule~5:~Age} \rightarrow \mathtt{GoP} & \\
        &\mathtt{Rule~6:~Vocab~Diff} \rightarrow \mathtt{GoP} & \\
        &\mathtt{Rule~7:~SNR} \rightarrow \{\mathtt{Subs,~Del,~Ins~Error}\} & \\
        &\mathtt{Rule~8:~No.~Words} \rightarrow \{\mathtt{Subs,~Del,~Ins~Error}\} &
        \end{flalign*}
    \end{equation*}
    \caption{}
    \label{fig:rules}
\end{subfigure}
\vspace{0.2cm}
    \begin{subfigure}[t]{0.49\textwidth}
        \centering
        \includegraphics[width=0.9\textwidth]{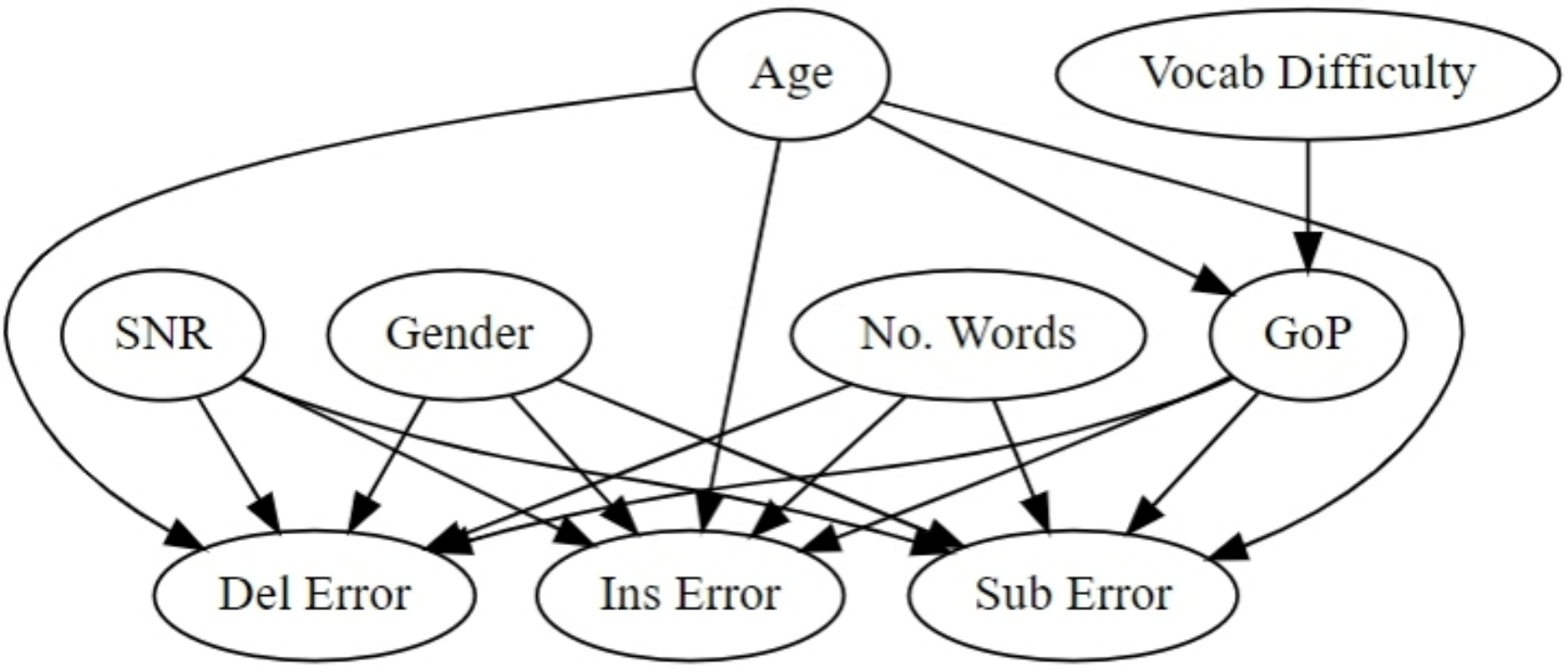}
        \vspace{0.5cm}
        \caption{}
        \label{fig:dag-our}
    \end{subfigure}
    \hfill
    \begin{subfigure}[t]{0.49\textwidth}
        \centering
        \vspace{-2cm}
        \begin{equation*}
        \begin{gathered}
        \mathtt{p}_{\mathtt{child}} = \mathtt{p}(\mathtt{Age}) \times \mathtt{p}(\mathtt{Gender}) \times \mathtt{p}(\mathtt{SNR})  \\
        \times \mathtt{p}(\mathtt{GoP} \mid \mathtt{Vocab~Difficulty}) \\
        \times \mathtt{p}(\mathtt{Vocab~Difficulty}) \times \mathtt{p}(\mathtt{No.~Words}) \times \mathtt{p}(\mathtt{GoP} \mid \mathtt{Age}) \\
        \times \mathtt{p}(\mathtt{Ins~Error} \mid \mathtt{Age,~Gender,~SNR,~GoP,~No.~Words}) \\
        \times \mathtt{p}(\mathtt{Del~Error} \mid \mathtt{Age,~Gender,~SNR,~GoP,~No.~Words}) \\
        \times \mathtt{p}(\mathtt{Subs~Error} \mid \mathtt{Age,~Gender,~SNR,~GoP,~No.~Words})
        \end{gathered}
        \end{equation*}
        \caption{}
        \label{fig:equation}
    \end{subfigure}
    
    \caption{Proposed causal model for explaining substitution (subs), deletion (del), and insertion (ins) errors in children's automatic speech recognition. (a.) Categorization of variables used in Causal graph. In node type, the fork indicates the nodes that have multiple children. A collider is a node is influenced by two or more nodes. A confounder is a node that influences both the dependent variable and independent variable, causing a spurious association. (b.) Factors affecting children's speech production, reconstructed based on finding in~\cite{workforce_children}. (c.) Rules for Causal graph construction. (d.) Proposed Causal graph for explainable children's ASR based on given rules in sub figure~\ref{fig:rules}. (e.) Joint probability distribution of physiology, cognitive response, extrinsic factors, and ASR errors for given Causal graph in sub figure~\ref{fig:dag-our}, which forms the basis for CI \cite{causal-quantification}.}
    \label{fig:combined-ci}
\end{figure}

\begin{figure}[htb]
\centering
\begin{minipage}{0.48\textwidth}
    \centering
    \includegraphics[width=\textwidth]{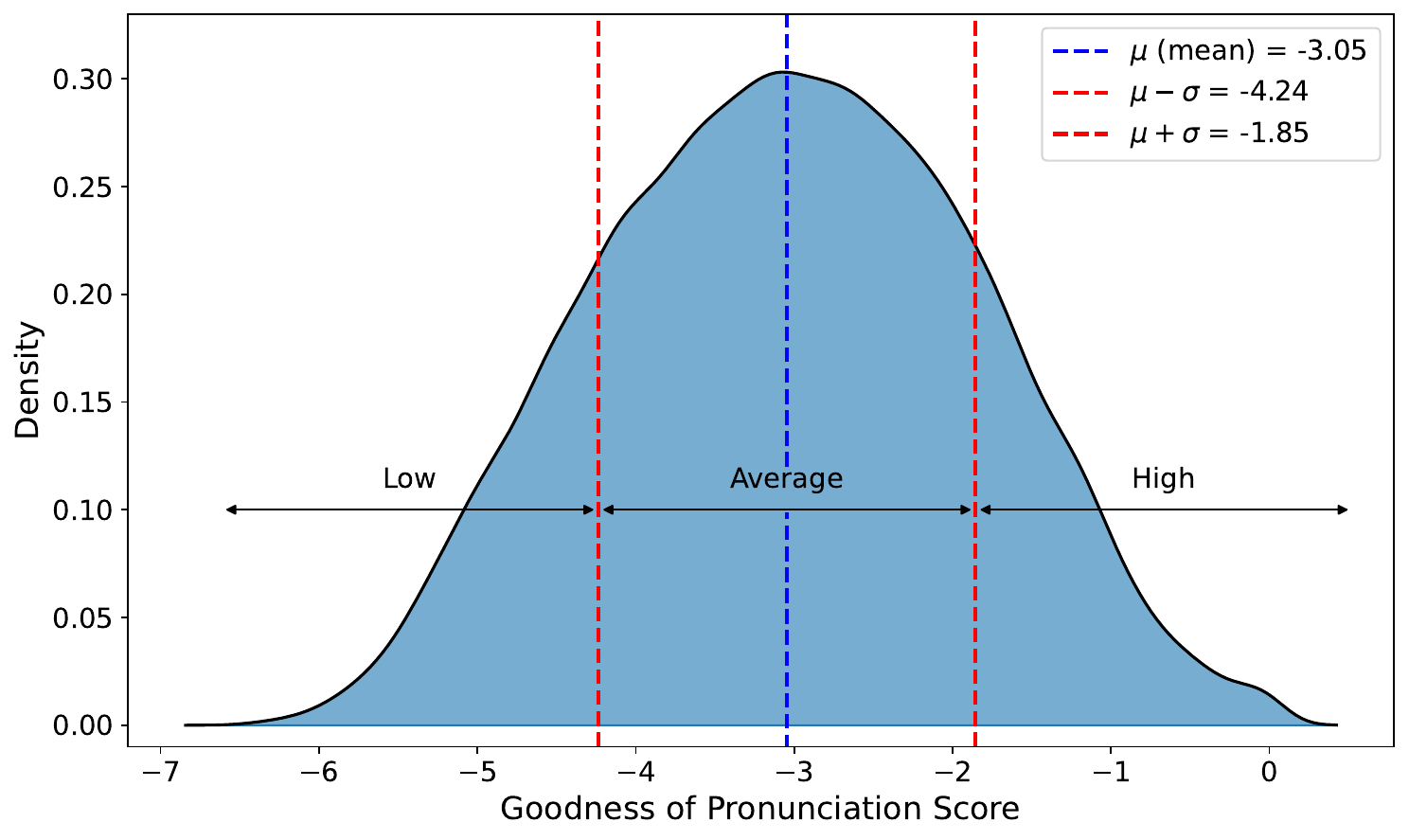}
    \subcaption{}
    \label{fig:gop}
\end{minipage}\hfill
\begin{minipage}{0.48\textwidth}
    \centering
    \includegraphics[width=\textwidth]{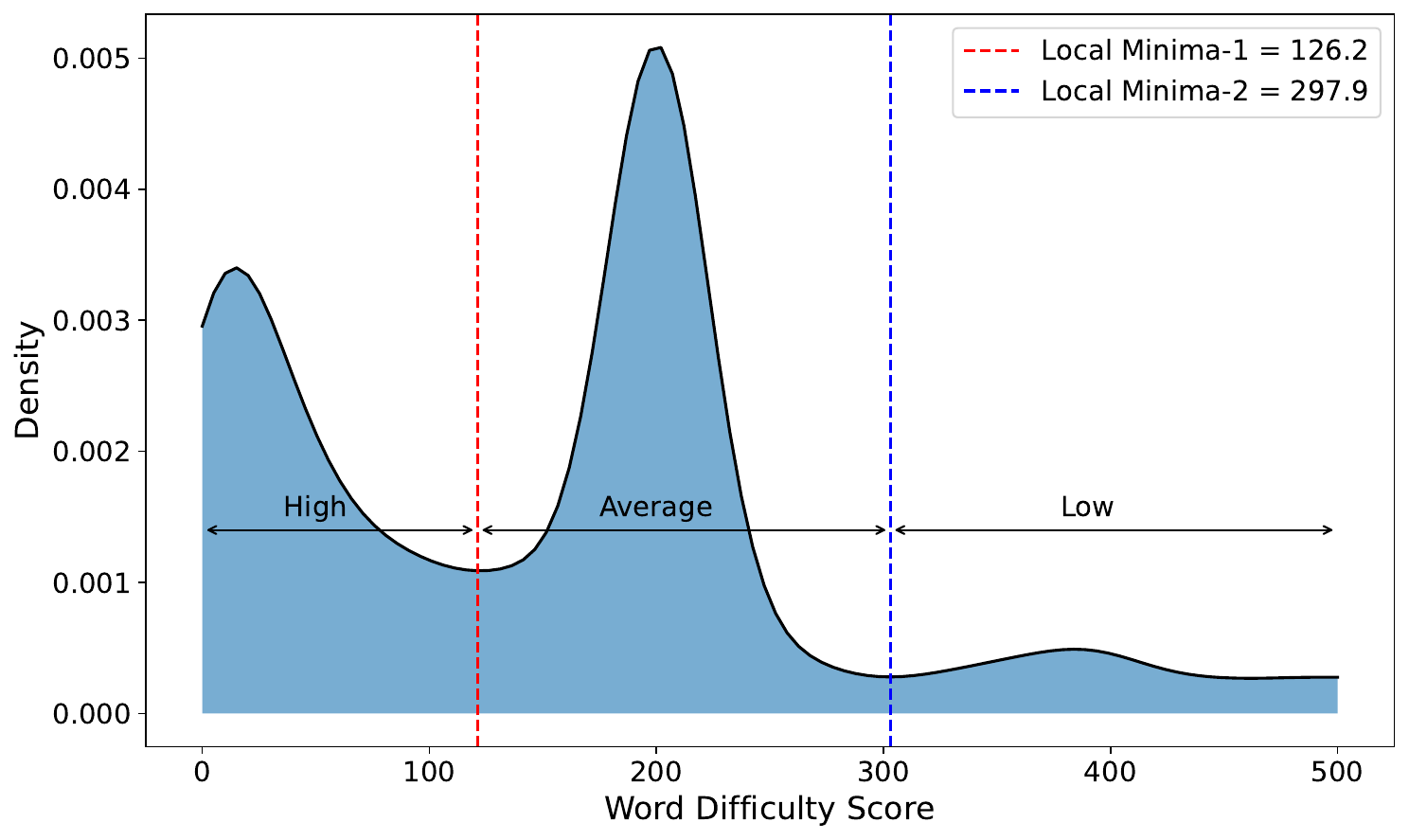}
    \subcaption{}
    \label{fig:wordif}
\end{minipage}
\caption{(a)~The distribution of the Goodness of Pronunciation (GoP) scores in the CSLU Kids corpus, categorized into three levels: High, Average, and Low. The categorization is based on the GoP score's distance from the mean, where scores within \( \pm \sigma \) (standard deviation) are classified as "Average," while scores beyond this range are labeled as "Low" or "High."; (b)~Distribution of word difficulty scores in the CSLU Kids corpus, categorized as "Low", "Average", and "High". The categorization of vocabulary difficulty scores is based on local minima in the kernel density estimates of the CSLU Kids corpus.}
\label{fig:combined}
\end{figure}

\section{Causal Modeling}\label{case2}

The results presented thus far indicate substantially high ASR error rates for children, particularly in the younger age groups. While our analysis goes beyond WER and includes two corpora in addition to an extensive set of four modern ASR models, our results primarily confirm (and extend) similar findings reported in prior studies. We now take one further step towards understanding the role of the various factors that contribute to the error rates, through causal post-hoc modeling motivated in Section \ref{sec:relwork}. The rationale for the proposed cause-effect analysis emerges from 
studies that indicates that speech production by children is impacted by a number of hidden factors including cognitive and physiological factors~\cite{workforce_children,cognitive_linguistic}. Our causal approach provides a principled way to quantify these effects. 

\subsection{Causal Structure for Explainable Children's ASR}
\label{method}

The proposed 
causal model 
for explaining the causes of degradation in children's ASR is shown in Figure~\ref{fig:combined-ci}. The high-level, abstract view of the model (see Figure ~\ref{fig:speech}) is inspired from the findings reported in ~\cite{workforce_children}. We assume the observed acoustic realization of speech produced by children to depend on three broad categories of factors---\textbf{physiological}, \textbf{cognitive}, and \textbf{extrinsic} ones. The first relates to the anatomical constraints (the physical speech production system) and the second to the cognitive constraints that influence speech production. Finally, the 
extrinsic factors relate to the environment and linguistic factors; 
they do not depend on the particular individual, but are nonetheless expected to influence ASR performance.

Each of the three broad categories of factors are characterized through specific sets of random variables, displayed in Fig.~\ref{fig:table-var}. Some of these variables are available in the corpus metadata (directly observed), whereas others need to be extracted (inferred) from either the audio or text data. The variables listed in Table~\ref{fig:table-var} are assumed to influence the three types of errors (for a given ASR system) either directly or indirectly. The assumed cause-effect relationships between the included variables can be described in a number of alternative, equivalent ways: Fig. \ref{fig:combined-ci} shows (c) a list of cause $\rightarrow$ effect rules; (d) a directed acyclic graph (DAG) where the direction of arrows points similarly from cause to effect; and (e) a factorized joint probability distribution.

Following~\cite{cognitive_linguistic,workforce_children}, we characterize the physiological response through age and gender, and cognitive response through so-called \emph{goodness of pronunciation} (GoP) score~\cite{gop}, which will be 
detailed below. 
The extrinsic factors include background noise, vocabulary difficulty~\cite{word_difficulty}, and the number of words in the audio. 
While age and gender metadata are given, the other variables indicated need to be inferred. Besides age, GoP is assumed to be impacted by the vocabulary difficulty (detailed below). This is based on the premise that children learn the easy and common words first. Additionally, extrinsic factors, including background noise (measured through estimated signal-to-noise ratio or SNR), and the number of words in context, impacts ASR error rates.

\subsection{Experimental Setup}
\label{ci-exp}
In the first part of the analysis, we analyze the sensitivity 
of the open-source SFMs 
to children's physiology (characterized by age and gender~\cite{workforce_children}), cognition (characterized by pronunciation ability~\cite{cognitive_linguistic}), and extrinsic factors. 
In the second part, we repeat the analysis with fine-tuned SFMs. 

Let us first detail how the variables requiring inference are computed in practice, starting with GoP. 
Various confidence measures~\cite{confidence_measures} have been proposed for automatic pronunciation quality assessment. This includes likelihood ratio~\cite{gop1} and posterior probability~\cite{gop2} of target phonemes 
obtained through forced alignment. The 
GoP score~\cite{gop} is a variation of posterior probability, which can be used for pronunciation assessment in~\cite{gop3}. It 
normalizes the posterior probability of the target phoneme 
the maximum posterior probability across all possible phonemes, 
as follows:
\[
GOP(p) \approx \log \frac{P(p|o_t; t_s, t_e)}{\max_{q \in Q} P(q|o_t; t_s, t_e)},
\]
where \( p \) represents the target phone and 
$Q$ indicates the set of all possible phones in the vocabulary. Here, the log posterior of phone \( p \) is computed as,
\[
\log P(p|o_t; t_s, t_e) \approx \frac{1}{t_e - t_s} \sum_{t_s}^{t_e} \log \sum_{s \in p} P(s|o_t),
\]
where \( t_s \) and \( t_e \) are the start and end indices of the phone (obtained using forced alignment); \( s \in p \) is the set of context-dependent phones; \( o_t \) is the input feature at frame \( t \); and \( P(s|o_t) \) is the frame-level posterior score 
obtained by aligning the utterance with the corresponding target phone sequence;
we use the \emph{Montreal forced aligner} (MFA) \cite{mfa} tool. 

The average of per-phoneme GoP scores is used 
as the utterance level pronunciation quality score. For 
causal inference, 
we further discretize the utterance-level score into three ordinal categories, \emph{Low}, \emph{Average}, and \emph{High}. 
This allows us to stratify the data into meaningful groups, which helps in identifying cause-effect relationships more precisely. By analyzing each category separately, we can understand how different levels of pronunciation quality influence the model's performance, and identify specific factors that contribute to degraded outcomes. The categorization of the GoP-based pronunciation score in the CSLU Kids corpus is depicted in Fig.\ref{fig:gop}.

As for the word difficulty, certain words can be considered \emph{a priori} more complex to produce, whether due to simply being rare, or otherwise expected to involve a complex sequence of articulatory gestures. Since 
\emph{vocabulary difficulty} 
measures---
computed solely from text (prompts)---are dependent neither on the individual, nor the acoustic realization, it is of great interest to consider their role in the analysis of ASR errors.
To determine word difficulty, we employ a \emph{rarity} score based approach. Following~\cite{word_difficulty}, rarity can be measured by the inverse of the word’s frequency across various text corpora~\cite{word_difficulty}. To this end, we include the following 6 datasets, \emph{movie-reviews} (movie sentiments), \emph{Reuters} (news articles), \emph{brown} (American English texts), \emph{Gutenberg} (classic literature), \emph{webtext} (informal web content), and \emph{nps-chat} (online chat conversations). Words that are less frequent in 
these corpora posses 
higher rarity scores and are 
potentially more difficult to speakers \cite{infant_vocab}. 
To obtain a sentence-level difficulty score, 
we average the difficulty scores of the individual words. Similar to GoP, for causal inference 
we discretize the score to 
three difficulty categories: \emph{Low}, \emph{Average}, and \emph{High}. 
The categorization of vocabulary difficulty scores is based on finding local minima in the kernel density estimates, on the CSLU Kids corpus, as visualized in Fig.~\ref{fig:wordif}.

\subsection{Results and Discussion}
\label{sec:rd}
\begin{table}[t]
\renewcommand{\arraystretch}{1.1}
\centering
\caption{Average causal effect analysis of different factors on substitution, deletion, and insertion errors on CSLU Test-good dataset.}
\footnotesize
\vspace{0.2cm}
\label{tab:res4}
\begin{tabular}{|c|c|ccc|ccc|}
\hline
\multirow{3}{*}{\textbf{Model}} & \multirow{3}{*}{\textbf{Cause}} & \multicolumn{3}{c|}{\textbf{Open-source}} & \multicolumn{3}{c|}{\textbf{Finetuned}} \\
\cline{3-8}
 & & \textbf{Subs} & \textbf{Del} & \textbf{Ins} & \textbf{Subs} & \textbf{Del} & \textbf{Ins} \\
\hline
\multirow{5}{*}{Wav2Vec} & Age & -4.36 & -0.10 & -1.60 & -2.12 & -0.04 & -2.81 \\
\cline{2-8}
 & Gender & 0.80 & 0.17 & 0.87 & 0.25 & 0.06 & 0.22 \\
\cline{2-8}
 & GoP & -1.07 & -0.25 & -0.88 & -0.63 & -0.14 & -0.54 \\
\cline{2-8}
 & SNR & -1.08 & 0.09 & 0.28 & -0.58 & 0.11 & 0.09 \\
\cline{2-8}
 & \# Words in Audio & -9.20 & 0.30 & -4.23 & -7.14 & 0.36 & -3.12 \\
\hline
\multirow{5}{*}{HuBERT} & Age & -4.10 & -0.06 & -2.99 & -- & -- & -- \\
\cline{2-8}
 & Gender & 0.86 & 0.02 & 1.04 & -- & -- & -- \\
\cline{2-8}
 & GoP & -1.51 & -0.20 & -1.45 & -- & -- & -- \\
\cline{2-8}
 & SNR & -0.99 & 0.091 & 1.42 & -- & -- & -- \\
\cline{2-8}
 & \# Words in Audio & -7.96 & 0.27 & -5.31 & -- & -- & -- \\
\hline
\multirow{5}{*}{Whisper} & Age & -3.25 & -0.20 & -2.16 & -3.0 & 0.32 & -2.22 \\
\cline{2-8}
 & Gender & 0.62 & 0.37 & 1.15 & 1.90 & 0.10 & 1.33 \\
\cline{2-8}
 & GoP & -1.25 & -0.18 & 0.35 & -1.10 &-0.49 & -0.05 \\
\cline{2-8}
 & SNR & -1.21 & 0.14 & -0.21 & -1.02 & 0.15 & -0.36 \\
\cline{2-8}
 & \# Words in Audio & -5.10 & 1.02 & -3.45 & -5.53 & 1.20 & -3.59 \\
\hline
\multirow{5}{*}{MMS} & Age & -3.60 & -0.06 & -5.89 & -- & -- & -- \\
\cline{2-8}
 & Gender & 0.66 & -0.17 & 1.96 & -- & -- & -- \\
\cline{2-8}
 & GoP & -1.51 & -0.33 & 3.78 & -- & -- & -- \\
\cline{2-8}
 & SNR & -2.19 & 0.06 & -7.86 & -- & -- & -- \\
\cline{2-8}
 & \# Words in Audio & -12.54 & 0.52 & -13.22 & -- & -- & --\\
\hline
\end{tabular}
\end{table}

\begin{figure*}[!ht]
    \centering
    \begin{subfigure}[b]{0.75\textwidth}
        \includegraphics[width=\textwidth]{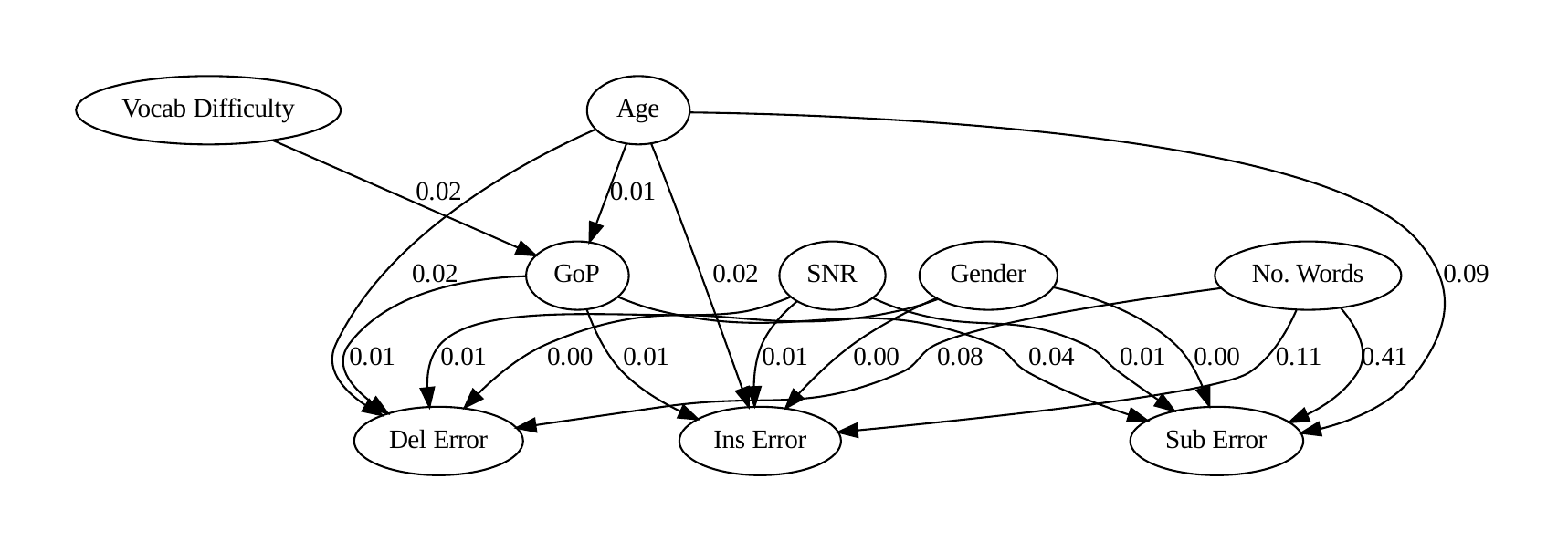}
        \caption{Wav2Vec2.0.}
        \label{fig:dag-w2v}
    \end{subfigure}

    \begin{subfigure}[b]{0.75\textwidth}
        \includegraphics[width=\textwidth]{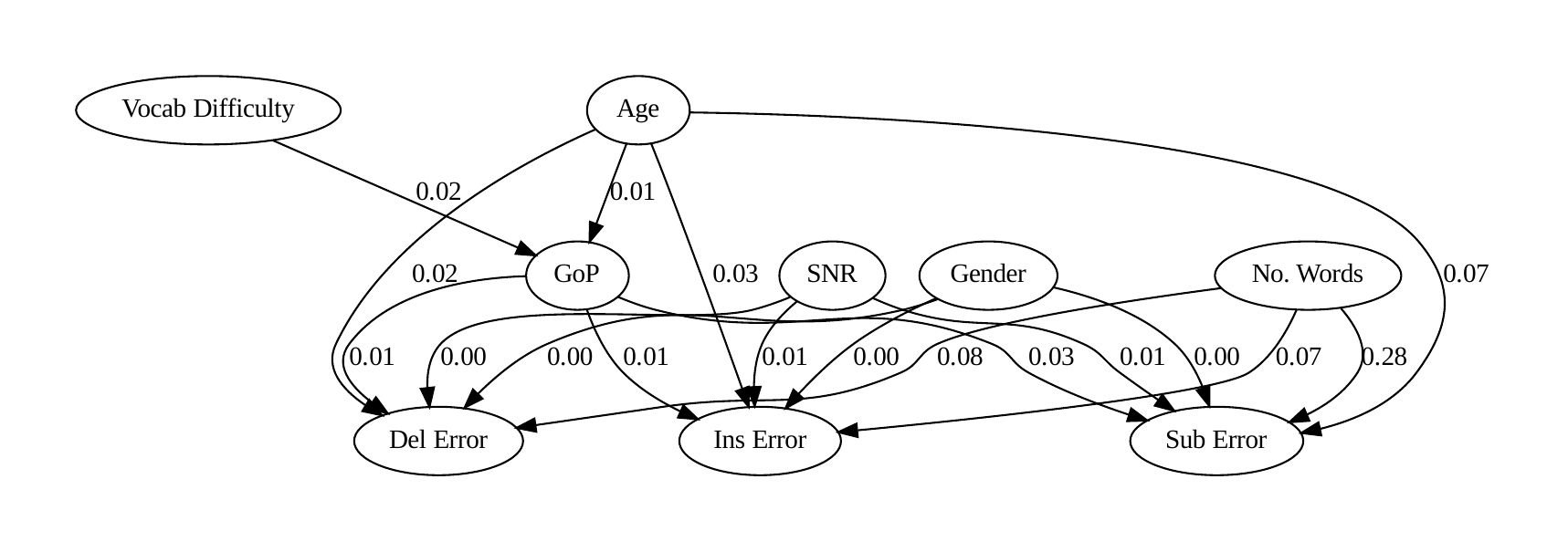}
        \caption{HuBERT.}
        \label{fig:dag-hubert}
    \end{subfigure}

    \begin{subfigure}[b]{0.75\textwidth}
        \includegraphics[width=\textwidth]{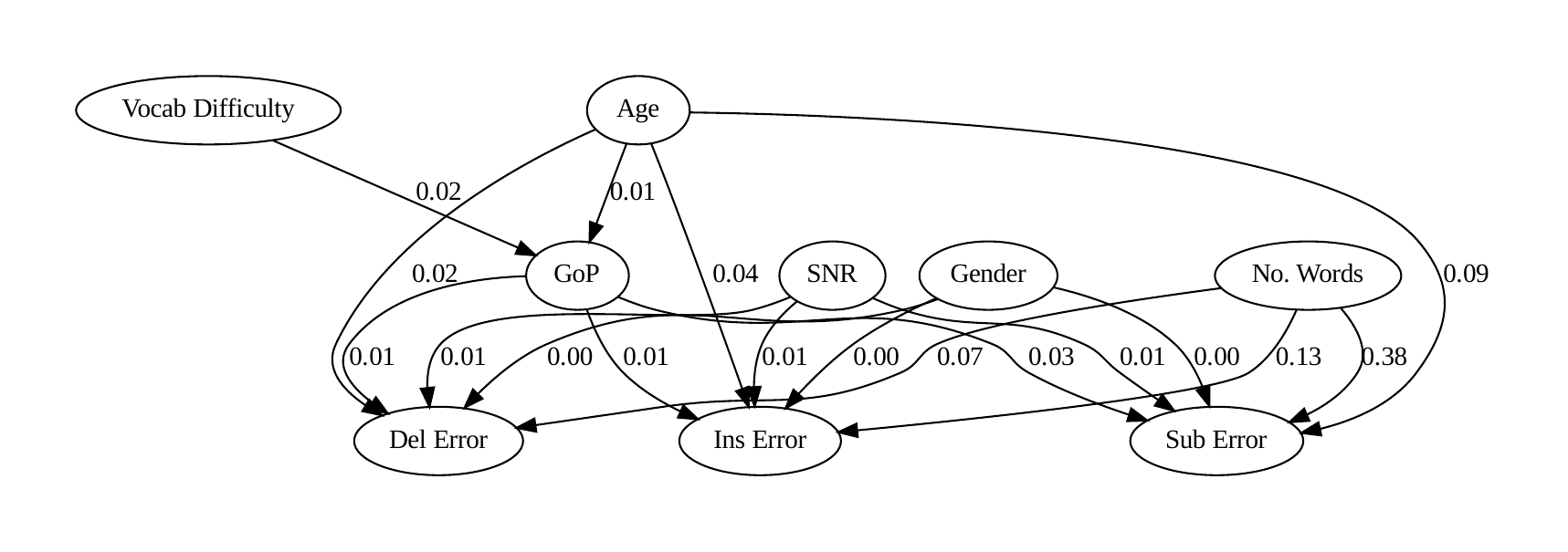}
        \caption{Whisper.}
        \label{fig:dag-whisper}
    \end{subfigure}
    \begin{subfigure}[b]{0.75\textwidth}
        \includegraphics[width=\textwidth]{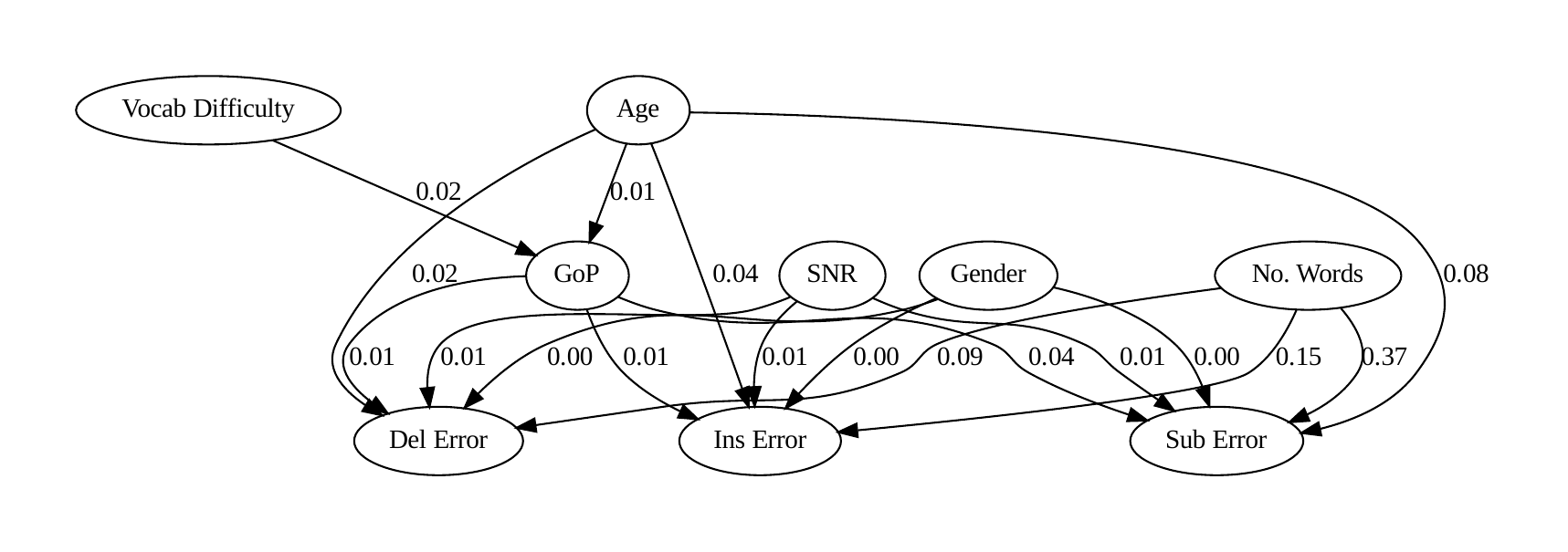}
        \caption{MMS.}
        \label{fig:dag-mms}
    \end{subfigure}
    \caption{Conditional Mutual Information based cause and effect analysis for different open-source models.}
    \label{fig:dag-os}
\end{figure*}

\begin{figure*}[t]
    \centering
    \begin{subfigure}[b]{0.75\textwidth}
        \includegraphics[width=0.9\textwidth]{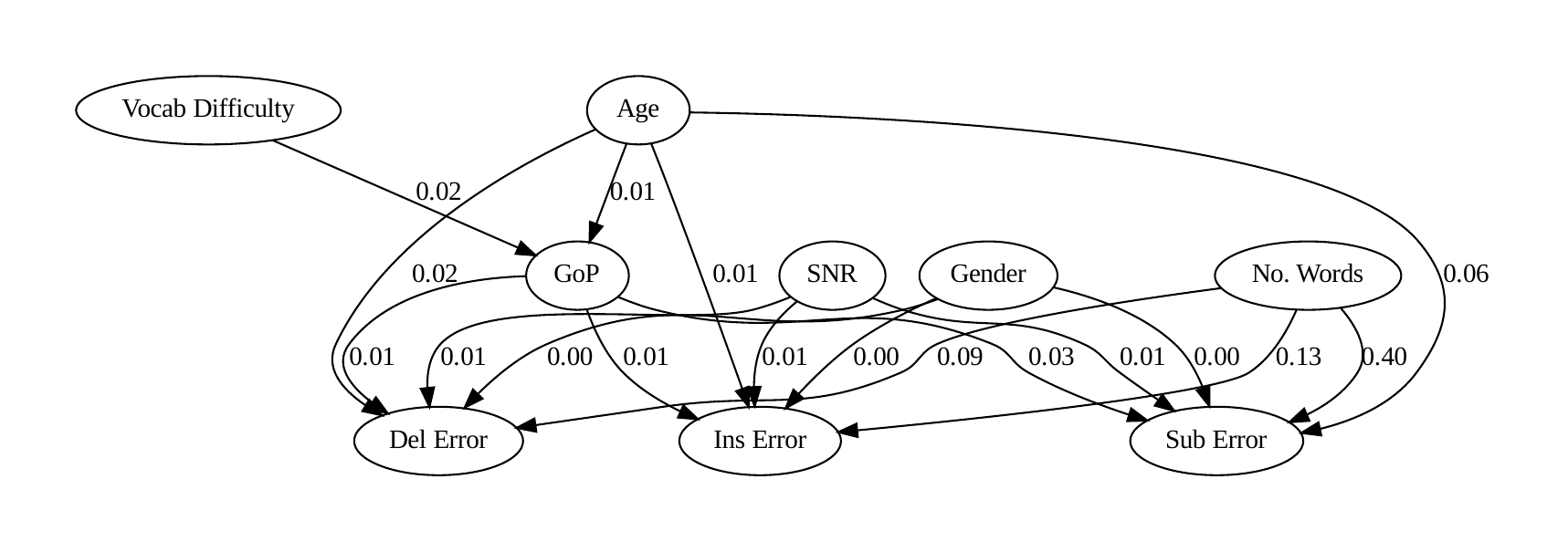}
        \caption{Wav2Vec2.0.}
        \label{fig:dag-w2v}
    \end{subfigure}
    \hfill
    \begin{subfigure}[b]{0.75\textwidth}
        \includegraphics[width=0.9\textwidth]{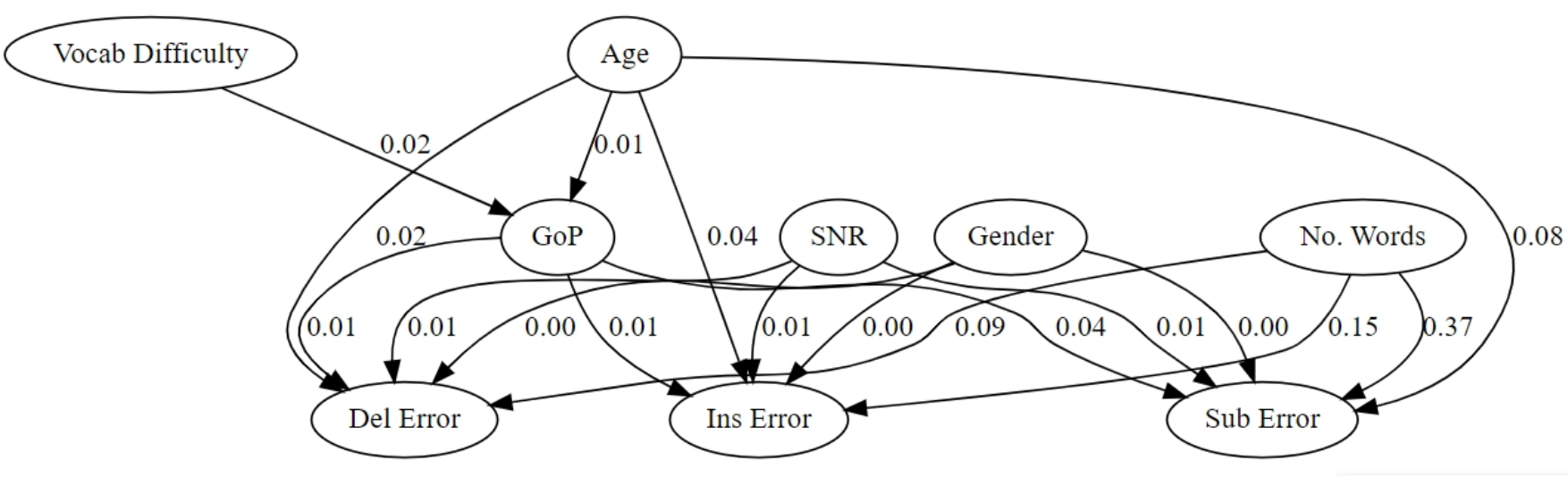}
        \caption{Whisper.}
        \label{fig:dag-whisper}
    \end{subfigure}
    \caption{Conditional Mutual Information based cause and effect analysis for different fine-tuned models.}
    \label{fig:dag-ft}
\end{figure*}


\label{sec:os-causal}
We first analyze the relative impact of the above causes on open-source speech foundation models through ACE values. The results are presented in Table~\ref{tab:res4}. 
A higher ACE of a particular cause indicates a higher impact of that cause on a particular error. 
Clearly, the 
number of words in audio and the age 
have the highest impact on all the three types of errors 
and for all the four ASR models. The negative sign 
for these two causes indicates that the errors decrease, as the age or the number of words increases. 
The impact of these two causes is strongest on substitution errors, 
and lowest on deletion errors. Out of the four models, \texttt{Whisper} 
possesses the lowest ACEs for these two causes, 
suggesting it to be 
best suited for children's ASR. Apart from the number of words and age, 
GoP and SNR are other major causes that influence the 
error rates, whereas gender has the lowest effect across the models. 

Additionally, we also present the causal DAG for different models in Fig.~\ref{fig:dag-os}. Unlike ACE, which is limited to assessing \emph{linear} association between the cause and effect, CMI 
quantifies \emph{non-linear} 
association between cause and effect~\cite{causal-quantification}. A higher CMI value on the edge of DAG indicates stronger impact of the parent on its child node(s). Confirming to the findings of ACE, CMI also 
points to 
highest dependence of the three types of errors 
on the number of words and age, followed by GoP and SNR. 
As before, \texttt{Whisper} 
shows the lowest CMI between age/number-of-words and ASR errors. This part of the analysis answers the first question in Section~\ref{intro}. 

\begin{itemize}
    \item \textbf{Q-1: What is the relative impact of physiological, cognitive, and external factors on the performance of open-source SFMs for children's speech?}
    \item \emph{Answer: Our results indicate that the SFMs have the highest sensitivity towards a particular extrinsic factor, the number of words in context. The second highest sensitivity is towards the physiological factor Age, while the other physiological factor Gender has negligible sensitivity. The cognitive factor has the least sensitivity among the physiological, cognitive, and extrinsic factors.}
\end{itemize}

Finally, we also present CMI between \( \mathtt{Age} \rightarrow \mathtt{GoP} \), and \( \mathtt{Vocab Diff} \rightarrow \mathtt{GoP} \) Fig.~\ref{fig:dag-os} and Fig.~\ref{fig:dag-ft}. These two edges in our Causal DAGs help us to understand the factor affecting the pronunciation ability in children. We observe that the CMI between \( \mathtt{Vocab Diff} \rightarrow \mathtt{GoP} \) is higher than that of \( \mathtt{Age} \rightarrow \mathtt{GoP} \), indicating vocabulary difficulty affects the pronunciation ability more than the Age. This part of the analysis answers the third question in Section~\ref{intro}. 

\begin{itemize}
    \item \textbf{Q-2: Pronunciation of children's impacts the WER~\cite{analysis_icslp}. But, what influences children’s pronunciation ability more: their age or the complexity of the vocabulary they use?}
    \item \emph{Answer: While both age and vocabulary difficulty affect pronunciation ability, vocabulary difficulty has a higher impact than that of age.}
\end{itemize}

\label{sec:ft-causal}
The average causal effect (ACE) of the fine-tuned \texttt{Wav2Vec2.0} and \texttt{Whisper} models is shown in Table~\ref{tab:res4}, with conditional mutual information (CMI) on the edges of the causal directed acyclic graphs (DAGs) in Fig.~\ref{fig:dag-ft}.

As per the ACE presented in Table~\ref{tab:res4} and Fig.~\ref{fig:dag-ft}, fine-tuning gives the largest reduction in ACE for the Age, which indicates that fine-tuning reduces the sensitivity of SFM towards children's physiology. While a moderate reduction in ACE and CMI for cognitive factor GoP is observed. This might be due to the mismatch in vocabulary between MyST train and the CSLU Test corpus, as GoP depends on vocabulary. On the other hand, the sensitivity of fine-tuned \texttt{Wav2Vec2.0} and \texttt{Whisper} remains high for a particular extrinsic factor, the number of words in context. This indicates the limitation of SFMs towards transcribing shorter contexts. This part of the analysis addresses the third question Section~\ref{intro}.
\begin{itemize}
    \item \textbf{Q-3: How does fine-tuning speech foundation models on child-specific speech datasets reduce the model's sensitivity to children's age, pronunciation ability, and extrinsic factors? Which aspect benefits the most from fine-tuning, and which remains largely unaffected?}
    \item \emph{Answer: Our, findings indicate that fine-tuning largely reduces the sensitivity of SFMs towards the physiological factor characterized by age. While the sensitivity towards extrinsic factor, shorter context, remains.}
\end{itemize}

\section{Discussion}
The primary contribution of this study is the development of a causal framework for the post-hoc explanation of children’s ASR, marking it the first such effort, to the best of our knowledge. With the help of the proposed framework, the findings obtained from our experiments can be categorized into two main parts: confirmative findings and novel findings. In the \textbf{confirmative findings} aspect, we demonstrated through ACE and CMI that WER shows a decreasing trend with respect to the prominent physiological factor---age---confirming that older children tend to have lower WERs. On the other hand, the other physiological factor, gender, was confirmed to have a negligible effect on WER. These findings align with the previous studies~\cite{child_patel,gurunath_csl,child_benchmark} conducted on the CSLU Kids Corpus through non-causal approaches. From a methodological perspective, the causal approach provides a principled way to define a model that encodes cause-effect assumptions between the variables included. Hence, even if our results concerning age and gender variables merely confirm the findings reported in prior studies, the approaches used for reaching this conclusion are entirely different---from \emph{statistical} associative approaches (past work) to \emph{causal} analysis (our work). The consistency of our findings with the past research strengthens the reliability of our causal analysis and confirms that physiological factors, particularly age, play a significant role in children's speech recognition accuracy.

As for the \textbf{novel findings} aspect, our proposed causal model 
provides a quantitative understanding of the cognitive factor, measured through Goodness of Pronunciation (GoP), and its impact on WER in children’s speech. With help of ACE in~\ref{tab:res4} and CMI in Fig~\ref{fig:dag-os}, we observe the ACE and CMI values corresponding to GoP are lower than that of physiological factors, however, comparable with other factors such as signal-to-noise ratio, indicating the importance of pronunciation on children's ASR. Further, the CMI in~\ref{fig:dag-os} indicates that pronunciation is more impacted by difficulty of target vocabulary than age.

Beyond physiology and cognition, our study highlights the influence of extrinsic factors, including vocabulary difficulty, SNR, and the number of spoken words, that collectively affect children’s WERs. Our framework also quantifies these extrinsic factors, offering a holistic view of their role in ASR performance. From the ACE and CMI values in~\ref{tab:res4} and CMI in Fig~\ref{fig:dag-os}, we observe that all the SFMs are extremely sensitive towards the number of words in audio, the 
SNR being another important factor. Most importantly, with the growing trend of fine-tuning self-supervised speech foundation models (SFMs) for children’s speech, our framework explains the differential impact of fine-tuning across factors, addressing which among physiological, cognitive, and extrinsic factors benefit most and which remain unaffected. 

Our findings suggest that ASR researchers, even those not employing causal inference, can consider incorporating variables such as vocabulary difficulty, background noise, number of spoken words, and GoP into their modeling and experimental designs. These variables offer actionable insights into performance trends and can guide the development of more robust and generalizable ASR systems for children. By emphasizing the stability and interpretability of these relationships, our study provides a foundation for future research to refine and extend these principles.

\section{Conclusion}
To recap our initial motivation: children are a growing group of speech technology users who, unfortunately, are not treated fairly by this technology. In this study we focused on ASR implemented through large-scale foundational models, a class of models that is otherwise influential in the speech field. Rather than proposing another ASR architecture, or proposing to collect another large-scale children dataset (which is not without stringent privacy and ethical considerations), we have capitalized the importance to understand what factors, for given datasets and ASR models, lead the models to fail transcribe children's speech correctly. Causal methodology provide a principled way to approach this problem. More specifically, 
we proposed a post-hoc causal analysis approach to quantify the sensitivity of SFMs to children's physiological, cognitive, and extrinsic factors, addressing the existing knowledge gap in explainability for children’s automatic speech recognition (ASR) systems. Unlike prior work that primarily examines the impact of age on WERs, our study incorporates cognitive factors (e.g., GoP) and extrinsic variables (e.g., vocabulary difficulty, signal-to-noise ratio) to provide a more detailed understanding of ASR errors in children. These findings open new avenues for future research, such as developing data augmentation techniques to enhance ASR robustness against mispronunciations. 

While our causal inference framework presents a step forward in explainable children's ASR, it is important to acknowledge its limitations. First, the choice of variables in our analysis and their relation established through DAG, though inspired by existing studies, remains somewhat subjective. It is good to point though that the choice of the variables to be included (and those to be excluded), as well as the choice of the causal model are eventually subjective (see, e.g. ~\cite{halpern})---the infamous 'domain expertise'. The choices of the variables and the causal model are also dictated by the purpose of a study (here, an attempt to explain ASR errors in a post-hoc way). The second limitation is that 
our findings are derived from a single dataset, limited to a specific language and demographic, which may constrain their generalizability. 
Even if, currently, there are limited data resources to address ASR in broader demographic populations, we deem these important additions. Apart from the gender and age variables addressed in our modeling, factors such as language could be incorporated to more complete causal models.

Looking ahead, we plan to expand this framework to address these limitations by incorporating causal structure discovery on variables and testing on multiple datasets. More importantly, although we apply our causal inference approach to children's ASR, it can be easily extended to other speech-processing tasks, such as assessing the sensitivity of speech systems to different speech attributes in spoofed speech detection, paving the way for explainable speech AI. We plan to share the analysis scripts and code via a public repository (e.g., GitHub) to facilitate future research and support reproducibility.
\appendix
\section{Causal Inference}
\label{app1}
Causal inference aims to establish cause-and-effect relationships, that go beyond mere statistical (correlational) associations~\cite{causal-survey, causal-quantification}. Concretely, causal relations are formalized through \emph{directed acyclic graphs} (DAGs). Given a joint distribution over \( n \) random variables, a DAG encodes all information about how interventions on one variable affect the others. Formally, let \( G \) be a causal DAG with nodes \( X_1, \ldots, X_n \) (one per each variable). An edge \( X_i \rightarrow X_j \) implies that \( X_i \) causally influences \( X_j \).

\begin{figure}[h]
    \centering
    \includegraphics[width=0.5\textwidth]{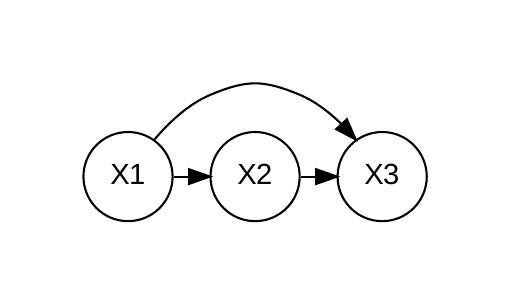}
    \caption{Directed Acyclic Graph (DAG) illustrating the causal relationships where $X_2$ is a parent of $X_3$, $X_1$ is a parent of both $X_2$ and $X_3$, making $X_1$ a confounder. $X_2$ and $X_3$ are descendants of $X_1$.}
    \label{fig:dag_example}
\end{figure}

In the example shown in Figure~\ref{fig:dag_example}, \(X_1\) has no parents,  \(X_2\) has \(X_1\) as its parent (denoted by \(\text{PA}_2 = \{X_1\}\)), \(X_3\) has \(X_2,X_1\) as its parent (\(\text{PA}_3 = \{X_2, X_1\}\)), and $X_2$ and $X_3$ are descendants of $X_1$. According to the factorization rule~\cite{bishop2024}, the joint probability distribution \(P(X_1, X_2, X_3)\) can be written as: \[ P(X_1, X_2, X_3) = P(X_1) \cdot P(X_2 \mid X_1) \cdot P(X_3 \mid X_2).\] This factorization reflects the structure of the DAG, where \(P(X_1)\) is the marginal probability of \(X_1\) since \(X_1\) has no parents. \(P(X_2 \mid X_1)\) is the conditional probability of \(X_2\) given \(X_1\). \(P(X_3 \mid X_2)\) is the conditional probability of \(X_3\) given \(X_2\). 

According to the so-called \emph{causal Markov condition} (CMC)~\cite{causal-survey, causal-quantification}, assumed for the models used in this study, every node \(X_j\) is conditionally independent of its non-descendants, given its parents with respect to the causal DAG. 
CMC is important in simplifying the analysis of complex systems. By identifying a \emph{Markov sufficient set} 
 (defined as the smallest set of variables that renders a node conditionally independent of all others, ensuring no information is lost in terms of causal dependencies), one can reduce the dimensionality of the problem and focus on a subset of variables that captures all the dependency information. By assuming CMC sufficiency, the joint probability distribution \( P(X_1, \ldots, X_n) \) factorizes according to the DAG structure as:
\begin{equation}
P(X_1, \ldots, X_n) = \prod_{j=1}^n P(X_j | \text{PA}_j),
\end{equation}
where \( \text{PA}_j \) denotes the set of parent variables of \( X_j \) in \( G \).


Once the causal DAG is constructed, the next task is to measure the strength of the causal association between nodes in the DAG. Two common approaches for measuring the strength of causal relation 
are (1) \emph{average causal effect} (ACE)~\cite{causal-quantification}, and (2) \emph{mutual information}.
The 
former measures the average difference in outcomes between two hypothetical scenarios: one where a particular variable (the “treatment”) is set to one value (e.g., 1) and another where it is set to an alternative value (e.g., 0). In our ASR context, `treatment' simply refers to the presence or absence of a specific factor under investigation (e.g., whether a child’s speech is influenced by a particular physiology, cognitive or extrinsic variable). 
For a binary treatment variable, ACE is defined as:
\begin{equation}
\text{ACE}(X_i \rightarrow X_j) = P(X_j = 1 | do(X_i = 1)) - P(X_j = 1 | do(X_i = 0))
\end{equation}

This allows us to quantify the causal impact of specific variables on ASR performance, helping identify which factors considerably influence outcomes such as word error rates.

The second measure of causal strength, mutual information~\cite{cover}, quantifies the amount of information that one random variable contains about another random variable. It quantifies the reduction in uncertainty about one variable given knowledge of the other. For two discrete random variables \( X \) and \( Y \), mutual information \( I(X; Y) \) is defined as:
\begin{equation}
I(X; Y) = \sum_{x \in X} \sum_{y \in Y} P(x, y) \log \left( \frac{P(x, y)}{P(x) P(y)} \right),
\end{equation}
where \( P(x, y) \) is the joint probability distribution of \( X \) and \( Y \), and \( P(x) \) and \( P(y) \) are the marginal probability distributions. Mutual information can alternatively be expressed as, 
\begin{equation}
I(X; Y) = H(X) - H(X | Y) = H(Y) - H(Y | X),
\end{equation}
where \( H(X) \) is the entropy of \( X \), and \( H(X | Y) \) is the conditional entropy of \( X \) given \( Y \). Entropy \( H(X) \) is defined as,
\begin{equation}
H(X) = - \sum_{x \in X} P(x) \log P(x),
\end{equation}
and the conditional entropy as, 
\begin{equation}
H(X | Y) = - \sum_{x \in X} \sum_{y \in Y} P(x, y) \log P(x | y).
\end{equation}
Mutual information is always non-negative and is zero if (and only if) \( X \) and \( Y \) are independent.

\end{document}